\documentclass[longauth,oneside]{aa}

\usepackage[varg]{txfonts}
\usepackage{natbib}
\usepackage{graphicx}
\usepackage{subcaption}
\usepackage{wasysym} 
\usepackage{rotating}
\usepackage{multirow} 
\usepackage{xcolor} 
\usepackage{amsmath}
\usepackage{ulem} 
\usepackage{blindtext}
\usepackage{color}
\usepackage[title,toc]{appendix}
\usepackage{threeparttable}
\usepackage{amssymb}
\usepackage{longtable}

\newcommand{\msun}{\mbox{M$_{\odot}$}}

\newcommand{\rsun}{\mbox{R$_{\odot}$\,}}
\newcommand{\kms}{\mbox{$\rm{km}\,s^{-1}$}}

\DeclareMathAlphabet{\mathsc}{OT1}{cmr}{m}{sc}
\def\testbx{bx}%
\DeclareRobustCommand{\ion}[2]{%
\relax\ifmmode
\ifx\testbx\f@series
{\mathbf{#1\,\mathsc{#2}}}\else
{\it{#1\,\mathsc{#2}}}\fi
\else\textup{#1\,{\mdseries\textsc{#2}}}%
\fi}
\newcommand{\Ha} {\mbox{H$\alpha$}\,}
\newcommand{\Hb} {\mbox{H$\beta$}\,}

\begin{document}

\title{Observations of the luminous red nova AT\,2021biy in the nearby galaxy NGC 4631\thanks{Photometric tables are only available in electronic form
at the CDS via anonymous ftp to \url{cdsarc.u-strasbg.fr (130.79.128.5)}
or via \url{http://cdsweb.u-strasbg.fr/cgi-bin/qcat?J/A+A/}}:}
\subtitle{Forbidden hugs in pandemic times -- III}

\author{
Y.-Z. Cai \inst{\ref{inst1} \thanks{E-mail: yzcai789@163.com}}
\and A. Pastorello\inst{\ref{inst2}} 
\and M. Fraser \inst{\ref{inst4}} 
\and X.-F. Wang\inst{\ref{inst1},\ref{inst5} \thanks{E-mail: wang\_xf@mail.tsinghua.edu.cn}}
\and A. V. Filippenko\inst{\ref{inst55}}
\and A. Reguitti\inst{\ref{inst2},\ref{inst19},\ref{inst20}}
\and K.~C.~Patra\inst{\ref{inst55},\ref{inst56}}
\and V.~P.~Goranskij\inst{\ref{inst116},\ref{inst115}}
\and E. A. Barsukova\inst{\ref{inst115}}
\and T. G. Brink\inst{\ref{inst55}}
\and N. Elias-Rosa\inst{\ref{inst2},\ref{inst6} }
\and H. F. Stevance \inst{\ref{inst1111},\ref{inst22} }  
\and W. Zheng\inst{\ref{inst55}}
\and Y. Yang\inst{\ref{inst55}}
\and K.~E.~Atapin\inst{\ref{inst116}}
\and S.~Benetti\inst{\ref{inst2}}
\and T. J. L. de Boer \inst{\ref{inst17}}
\and S. Bose\inst{\ref{inst35},\ref{inst36}}
\and J. Burke\inst{\ref{inst27},\ref{inst10}}
\and R.~Byrne\inst{\ref{inst4}} 
\and E. Cappellaro\inst{\ref{inst2}}
\and K. C. Chambers \inst{\ref{inst17}}
\and W.-L. Chen \inst{\ref{inst113}}
\and N.~Emami\inst{\ref{inst113}}
\and H. Gao \inst{\ref{inst17}}  
\and D.~Hiramatsu\inst{\ref{inst27},\ref{inst10},\ref{inst11},\ref{inst12}}
\and D. A. Howell\inst{\ref{inst27},\ref{inst10}}
\and M. E. Huber \inst{\ref{inst17}}
\and E. Kankare\inst{\ref{inst14},\ref{inst16}}
\and P.~L.~Kelly\inst{\ref{inst113}}
\and R.~Kotak\inst{\ref{inst14}}
\and T.~Kravtsov\inst{\ref{inst14}}
\and V.~Yu.~Lander\inst{\ref{inst116}}
\and Z.-T.~Li\inst{\ref{inst37},\ref{inst15}}
\and C.-C. Lin \inst{\ref{inst17}} 
\and P.~Lundqvist\inst{\ref{inst9}}
\and E. A. Magnier \inst{\ref{inst17}}
\and E.~A.~Malygin\inst{\ref{inst115}}
\and N.~A.~Maslennikova\inst{\ref{inst116},\ref{inst118}}
\and K.~Matilainen\inst{\ref{inst14}}
\and P.~A.~Mazzali\inst{\ref{inst41},\ref{inst25}} 
\and C.~McCully\inst{\ref{inst27}} 
\and J. Mo\inst{\ref{inst1}} 
\and S. Moran\inst{\ref{inst14}}
\and M.~Newsome\inst{\ref{inst27},\ref{inst10}}
\and D.~V.~Oparin\inst{\ref{inst115}}
\and E. Padilla Gonzalez\inst{\ref{inst27},\ref{inst10}}
\and T. M. Reynolds\inst{\ref{inst14},\ref{inst28}}
\and N.~I.~Shatsky\inst{\ref{inst116}}
\and S. J. Smartt \inst{\ref{inst22}}
\and K.~W.~Smith \inst{\ref{inst22}} 
\and M.~D.~Stritzinger\inst{\ref{inst23}}
\and A.~M.~Tatarnikov\inst{\ref{inst116},\ref{inst118}}
\and G.~Terreran\inst{\ref{inst27},\ref{inst10}}
\and R. I. Uklein\inst{\ref{inst115}}
\and G. Valerin\inst{\ref{inst2},\ref{inst3}}
\and P.~J.~Vallely\inst{\ref{inst35},\ref{inst36}}
\and O.~V.~Vozyakova\inst{\ref{inst116}}
\and R.~Wainscoat\inst{\ref{inst17}}
\and S.-Y. Yan\inst{\ref{inst1}}
\and J.-J. Zhang\inst{\ref{inst21},\ref{inst43}}
\and T.-M.~Zhang\inst{\ref{inst37},\ref{inst15}}
\and S.~G.~Zheltoukhov\inst{\ref{inst116},\ref{inst118}}
\and R.~Dastidar \inst{\ref{inst19},\ref{inst20}}
\and M.~Fulton \inst{\ref{inst22}}
\and L.~Galbany\inst{\ref{inst6},\ref{inst13}}
\and A.~Gangopadhyay \inst{\ref{inst7}}
\and H.-W.~Ge\inst{\ref{inst21},\ref{inst43},\ref{inst44}}
\and C.~P.~Guti\'errez\inst{\ref{inst14},\ref{inst31}}
\and H. Lin\inst{\ref{inst1}}
\and K.~Misra \inst{\ref{inst38}}
\and Z.-W. Ou \inst{\ref{inst39}}
\and I.~Salmaso\inst{\ref{inst2},\ref{inst3}}
\and L.~Tartaglia\inst{\ref{inst2}}
\and L.~Xiao \inst{\ref{inst111},\ref{inst112}}
\and X.-H.~Zhang\inst{\ref{inst1}}
}

\institute{
Physics Department and Tsinghua Center for Astrophysics (THCA), Tsinghua University, Beijing, 100084, China. \email{yzcai789@163.com} \label{inst1}
\and INAF - Osservatorio Astronomico di Padova, Vicolo dell'Osservatorio 5, 35122 Padova, Italy  \label{inst2}
\and School of Physics, O'Brien Centre for Science North, University College Dublin, Belfield, Dublin 4, Ireland  \label{inst4}
\and Beijing Planetarium, Beijing Academy of Science and Technology, Beijing, 100044, China. \email{wang\_xf@mail.tsinghua.edu.cn} \label{inst5}
\and Department of Astronomy, University of California, Berkeley, CA 94720-3411, USA \label{inst55}
\and Departamento de Ciencias Fisicas, Universidad Andres Bello, Fernandez Concha 700, Las Condes, Santiago, Chile \label{inst19}
\and Millennium Institute of Astrophysics (MAS), Nuncio Monsenor S\`{o}tero Sanz 100, Providencia, Santiago, 8320000 Chile \label{inst20}
\and Nagaraj-Noll-Otelline Graduate Fellow \label{inst56}
\and Sternberg Astronomical Institute of the Moscow State University, Moscow, 119234 Russia \label{inst116}
\and Special Astrophysical Observatory, Russian Academy of Sciences, Nizhnij Arkhyz, 369167 Russia \label{inst115}
\and Institute of Space Sciences (ICE, CSIC), Campus UAB, Carrer de Can Magrans, s/n, E-08193 Barcelona, Spain \label{inst6}
\and The Department of Physics, The University of Auckland, Private Bag 92019, Auckland, New Zealand \label{inst1111}
\and Astrophysics Research Centre, School of Mathematics and Physics, Queen’s University Belfast, Belfast BT7 1NN Northern Ireland, UK\label{inst22}
\and Institute for Astronomy, University of Hawaii, 2680 Woodlawn Drive, Honolulu HI 96822, USA \label{inst17}
\and Department of Astronomy, The Ohio State University, 140 W. 18th Avenue, Columbus, OH 43210, USA \label{inst35}
\and Center for Cosmology and AstroParticle Physics (CCAPP), The Ohio State University, 191 W.Woodruff Avenue, Columbus, OH 43210, USA \label{inst36}
\and Las Cumbres Observatory, 6740 Cortona Dr. Suite 102, Goleta, CA, 93117, USA \label{inst27}
\and Department of Physics, University of California, Santa Barbara, Santa Barbara, CA, 93106, USA \label{inst10}
\and Center for Astrophysics \textbar{} Harvard \& Smithsonian, 60 Garden Street, Cambridge, MA 02138-1516, USA \label{inst11}
\and The NSF AI Institute for Artificial Intelligence and Fundamental Interactions \label{inst12}
\and School of Physics and Astronomy, University of Minnesota, 116 Church Street SE, Minneapolis, MN 55455, USA \label{inst113}
\and Department of Physics and Astronomy, University of Turku, FI-20014 Turku, Finland \label{inst14}
\and Turku Collegium for Science, Medicine and Technology, University of Turku, FI-20014 Turku, Finland \label{inst16}
\and National Astronomical Observatories, Chinese Academy of Sciences, Beijing 100101, China \label{inst37}
\and School of Astronomy and Space Science, University of Chinese Academy of Sciences, Beijing 101408, China \label{inst15}
\and The Oskar Klein Centre, Department of Astronomy, Stockholm University, AlbaNova, SE-10691 Stockholm, Sweden \label{inst9}
\and Faculty of Physics, Moscow State University, 1 bldg. 2, Leninskie Gory, Moscow 119991, Russia \label{inst118}
\and Astrophysics Research Institute, Liverpool John Moores University, IC2, Liverpool Science Park, 146 Brownlow Hill, Liverpool L3 5RF, UK \label{inst41}
\and Max-Planck-Institut f\"ur Astrophysik, Karl-Schwarzschild Str. 1, D-85741 Garching, Germany\label{inst25}
\and Niels Bohr Institute, University of Copenhagen, Jagtvej 128, 2200 København N, Denmark \label{inst28}
\and Department of Physics and Astronomy, Aarhus University, Ny Munkegade 120, DK-8000 Aarhus C, Denmark \label{inst23}
\and Universit\`a degli Studi di Padova, Dipartimento di Fisica e Astronomia, Vicolo dell'Osservatorio 2, 35122 Padova, Italy \label{inst3}
\and Yunnan Observatories, Chinese Academy of Sciences, Kunming 650216, China \label{inst21}
\and Key Laboratory for the Structure and Evolution of Celestial Objects, Chinese Academy of Sciences, Kunming 650216, China \label{inst43}
\and Institut d’Estudis Espacials de Catalunya (IEEC), E-08034 Barcelona, Spain \label{inst13}
\and Hiroshima Astrophysical Science Centre, Hiroshima University, 1-3-1 Kagamiyama, Higashi-Hiroshima, Hiroshima 739-8526, Japan \label{inst7}
\and University of Chinese Academy of Sciences, Beijing 100049, People’s Republic of China \label{inst44}
\and Finnish Centre for Astronomy with ESO (FINCA), FI-20014 University of Turku, Finland \label{inst31}
\and Aryabhatta Research Institute of Observational Sciences, Manora Peak, Nainital 263 001, India \label{inst38}
\and School of Physics and Astronomy, Sun Yat-sen University, Zhuhai 519082, China \label{inst39}
\and Department of Physics, College of Physical Sciences and Technology, Hebei University, Baoding, 071002, China \label{inst111}
\and Key Laboratory of High-precision Computation and Application of Quantum Field Theory of Hebei Province, Hebei University, Baoding, 071002, China \label{inst112}
}

\date{Received Month XX, 2022 / Accepted Month XX, 2022}

\abstract{
We present an observational study of the luminous red nova (LRN) AT\,2021biy in the nearby galaxy NGC\,4631. The field of the object was routinely imaged during the pre-eruptive stage by synoptic surveys, but the transient was detected only at a few epochs from $\sim 231$\,days before maximum brightness. The LRN outburst was monitored with unprecedented cadence both photometrically and spectroscopically. AT\,2021biy shows a short-duration blue peak, with a bolometric luminosity of $\sim 1.6 \times 10^{41}$\,erg\,s$^{-1}$, followed by the longest plateau among LRNe to date, with a duration of 210\,days. A late-time hump in the light curve was also observed, possibly produced by a shell-shell collision. AT\,2021biy exhibits the typical spectral evolution of LRNe. Early-time spectra are characterised by a blue continuum and prominent H emission lines. Then, the continuum becomes redder, resembling that of a K-type star with a forest of metal absorption lines during the plateau phase. Finally, late-time spectra show a very red continuum ($T_{\mathrm{BB}} \approx 2050$ K) with molecular features (e.g. TiO) resembling those of M-type stars. Spectropolarimetric analysis indicates that AT\,2021biy has local dust properties similar to those of V838\,Mon in the Milky Way Galaxy. Inspection of archival {\it Hubble Space Telescope} data taken on 2003 August 3 reveals a $\sim 20$\,\msun\ progenitor candidate with log\,$(L/{\rm L}_{\odot}) = 5.0$\,dex and $T_{\rm{eff}} = 5900$\,K at solar metallicity. The above luminosity and colour match those of a luminous yellow supergiant. Most likely, this source is a close binary, with a 17--24\,\msun\ primary component. 
}
\authorrunning{Y.-Z. Cai et al.} 
\titlerunning{Luminous Red Nova AT\,2021biy}
\maketitle

\section{Introduction}\label{sec:intro} 

Modern wide-field surveys are discovering a large number of transients with intrinsic luminosity between that of core-collapse supernovae (CC~SNe) and classical novae (i.e. $-15 \lesssim M_V \lesssim -10$\,mag). These transients are often called `gap transients' \citep[e.g.][]{Kulkarni2009aaxo.conf..312K, Kasliwal2012PASA...29..482K, Pastorello2019NatAs...3..676P,Fraser2020RSOS....700467F}. They have heterogeneous observational properties, arise from different progenitors, and are likely triggered by diverse physical mechanisms. 

Luminous red novae \citep[LRNe; e.g.][]{Martini1999AJ....118.1034M,Munari2002A&A...389L..51M,Tylenda2005A&A...436.1009T,Ivanova2013A&ARv..21...59I,Kochanek2014MNRAS.443.1319K,Williams2015ApJ...805L..18W,Goranskij2016AstBu..71...82G,Pejcha2016MNRAS.461.2527P,Smith2016MNRAS.458..950S,Blagorodnova2017ApJ...834..107B,Lipunov2017MNRAS.470.2339L,MacLeod2017ApJ...835..282M,Pejcha2017ApJ...850...59P,Cai2019A&A...632L...6C,Pastorello2019A&A...625L...8P,Pastorello2019A&A...630A..75P,Stritzinger2020A&A...639A.104S, Pastorello2021A&A...646A.119P,Pastorello2021A&A...647A..93P,Blagorodnova2021A&A...653A.134B} are a subclass of gap transients showing well-constrained observational features. They usually have a composite light curve \citep[e.g.][]{Kankare2015A&A...581L...4K} and dramatic spectral evolution. In particular, LRNe show a long-lasting, slow luminosity rise before the outburst, followed by a double-peaked light curve, with the second peak sometimes resembling a sort of plateau.
Their early-time optical spectra are similar to those of Type IIn supernovae \citep[SNe\,IIn;][]{Schlegel1990MNRAS.244..269S, Filippenko1997ARA&A..35..309F} and are characterised by a blue continuum with superposed narrow H emission lines. In contrast, spectra taken during the second peak exhibit a redder continuum and a forest of metal lines in absorption, along with much weaker Balmer lines. At late times, spectra of LRNe transition to those of M-type stars, showing broad molecular absorption bands \citep[e.g. TiO; see][]{Kaminsk2009ApJS..182...33K, Mason2010A&A...516A.108M, Barsukova2014AstBu..69...67B, Cai2019A&A...632L...6C,Pastorello2021A&A...646A.119P,Pastorello2021A&A...647A..93P}. 

These well-defined observables allow us to discriminate LRNe from another subclass of gap transients, the so-called intermediate-luminosity red transients \citep[ILRTs; e.g.][Valerin et al. in prep.]{Botticella2009MNRAS.398.1041B, Thompson2009ApJ...705.1364T, Berger2009ApJ...699.1850B,Cai2018MNRAS.480.3424C, Cai2021A&A...654A.157C, Stritzinger2020A&A...639A.103S}. ILRTs display single-peak light curves, with slowly-evolving spectra dominated by H and Ca features. 
ILRTs are proposed to be electron-capture SN explosions \citep[e.g.][]{Nomoto1984ApJ...277..791N, Nomoto1987ApJ...322..206N, Poelarends2008ApJ...675..614P, Pumo2009ApJ...705L.138P, Moriya2014A&A...569A..57M, Doherty2015MNRAS.446.2599D, Doherty2017PASA...34...56D, Leung2020ApJ...889...34L} from super-asymptotic giant branch (SAGB) stars \citep[e.g.][]{Prieto2008ApJ...681L...9P, Prieto2009ApJ...705.1425P, Thompson2009ApJ...705.1364T, Botticella2009MNRAS.398.1041B, Adams2016MNRAS.460.1645A, Cai2018MNRAS.480.3424C, Cai2021A&A...654A.157C}. Nonetheless, controversial objects are occasionally observed, such as M85-2006OT1 and AT\,2018hso, sharing transitional properties with both LRNe and ILRTs \citep[][]{Kulkarni2007Natur.447..458K, Pastorello2007Natur.449E...1P, Rau2007ApJ...659.1536R,Cai2019A&A...632L...6C}. 

The LRN phenomenon can be interpreted as the result of a common-envelope ejection, likely followed by a stellar merging in a close binary system \citep[e.g.][]{Soker2003ApJ...582L.105S, Soker2006MNRAS.373..733S, Tylenda2011A&A...528A.114T, Ivanova2013A&ARv..21...59I, Soker2016MNRAS.462..217S, Metzger2017MNRAS.471.3200M, Mauerhan2018MNRAS.473.3765M, Segev2019ApJ...884...58S,Soker2020ApJ...893...20S, Soker2021RAA....21..112S}. 
The progenitor systems of LRNe span a wide range of masses, from massive binaries of a few tens of \msun, to low-mass systems even down to $\sim 1$\,\msun. By studying a sample of LRNe, a correlation between their luminosity and the mass of the progenitor systems has been established, with higher-mass progenitors producing more-luminous outbursts \citep[e.g.][]{Kochanek2014MNRAS.443.1319K,Blagorodnova2021A&A...653A.134B}.

Although ongoing surveys have significantly increased the available LRN sample in recent years, only a limited number of events have good data covering all evolutionary stages. In particular, pre-outburst detections, a high-cadence light curve, and well-sampled spectra are fundamental to properly compare LRNe with theoretical models. This motivated us to conduct an aggressive follow-up campaign for a new object, AT\,2021biy, discovered in the nearby galaxy NGC\,4631.

In this paper, we report the results of our observations of AT\,2021biy in the optical and near-infrared (NIR) domains, and our analysis of the progenitor and its environment from pre-discovery archival images. The paper is structured as follows. The discovery, distance, reddening, and host-galaxy properties are reported in Sect. \ref{sec:info}, while the photometric and spectroscopic evolution are respectively illustrated in Sect. \ref{sec:photometry} and \ref{sec:spectra}. Sect. \ref{progenitor} investigates the quiescent progenitor system of AT\,2021biy. A discussion and our conclusions are presented in Sect. \ref{sec:discussion}.

\section{Discovery, distance, reddening, and host galaxy} \label{sec:info}

\begin{figure}[htb]
\centering
\includegraphics[width=9.0cm]{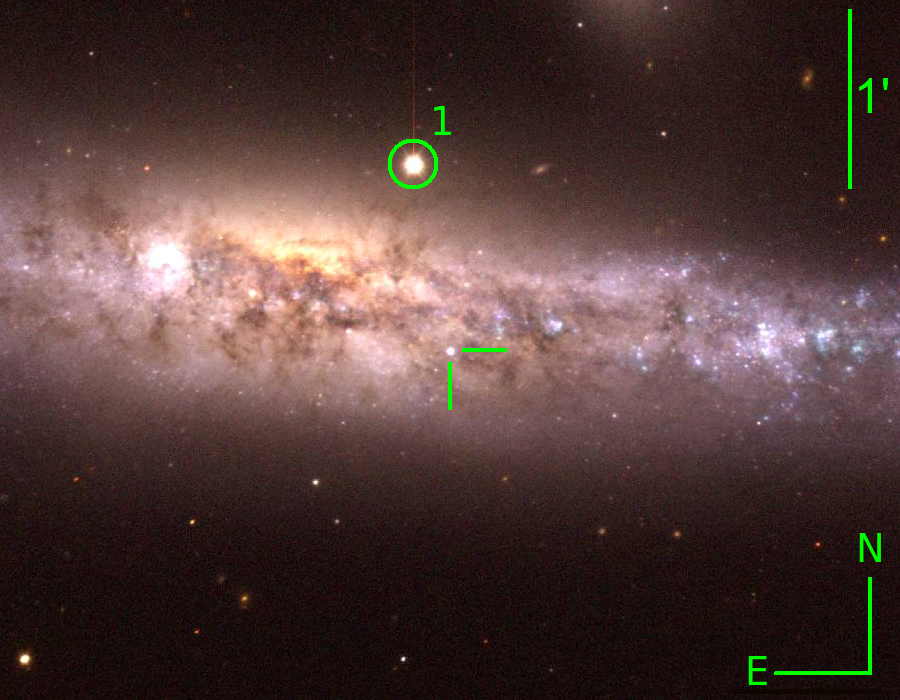}
\caption{Location of the LRN AT\,2021biy within the host galaxy NGC\,4631 in a colour $B+V+r$ image. 
The original filtered images were taken with NOT+ALFOSC on 10 February 2021.
AT\,2021biy is located 60.4\arcsec S and 13.2\arcsec W from the bright Star 1.}
\label{FC}
\end{figure}

AT\,2021biy\footnote{The object also has the survey designations ATLAS21dfy and PS21aaa.} was discovered by the Asteroid Terrestrial-impact Last Alert System \citep[ATLAS;][]{Tonry2018ApJ...867..105T,Tonry2018PASP..130f4505T, Smith2020PASP..132h5002S} on 2021 January 29.56 (UT dates are used throughout this paper; that epoch corresponds to MJD = 59243.56), at an ATLAS orange-filter ($o$) brightness $o=18.12\pm0.14$\,mag \citep{Smith2021TNSAN..33....1S,Tonry2021TNSTR.273....1T}. Soon after its discovery, it was classified as an LRN \citep{Cai2021ATel14360....1C} in the framework of the Nordic-optical-telescope Unbiased Transient Survey 2 (NUTS2\footnote{\url{http://nuts2.sn.ie/}}) collaboration. Its J2000 coordinates are $\alpha = 12^{\rm h} 42^{\rm m} 04.025^{\rm s}$, $\delta = +32\degr 32\arcmin 07.88\arcsec$, placing it $21.6\arcsec$ south and $50.4\arcsec$ west of the core of the SBd galaxy NGC\,4631. The location of the transient within the host galaxy is shown in Figure~\ref{FC}, while the main properties of NGC\,4631 are reported in Table \ref{hgalaxy}.

\begin{table}[htb]
\centering
\caption{General properties of NGC\,4631.}
\label{hgalaxy}
\begin{threeparttable}[b]
\begin{tabular}{lr}
\hline\hline
$\alpha$ (J2000)          & $12^{\rm h} 42^{\rm m} 08.0^{\rm s}$ \\
$\delta$ (J2000)         & $+32\degr32\arcmin29.4\arcsec$ \\
Morphological Type\tnote{1} & SB(s)d edge-on \\
$B$ Magnitude\tnote{2}  & 9.75 $\pm$ 0.16\,mag \\
Redshift\tnote{3}   & 0.002035 $\pm$ 0.000007 \\
Distance Modulus\tnote{4}	& 29.36 $\pm 0.15$\,mag\\
MW extinction ($A_{V,{\rm MW}}$)\tnote{5}  & 0.047\,mag \\
12 + log(O/H) ($R=0$)\tnote{6} & 8.39 $\pm$ 0.06\,dex\\
\hline\hline
\end{tabular}
\medskip
References: 1 = \citet{deVaucouleurs1991rc3..book.....D}; 2 = \citet{Kennicutt2008ApJS..178..247K}; 3 = \citet{Wolfinger2013MNRAS.428.1790W}; 4 = \citet{Monachesi2016MNRAS.457.1419M}; 5 = \citet{Schlafly2011ApJ...737..103S}; 6 = \citet{Pilyugin2004A&A...425..849P, Pilyugin2014AJ....147..131P}.
\end{threeparttable}
\end{table}

The adopted distance of NGC\,4631 is $d = 7.46 \pm 0.50$\,Mpc ($\mu = 29.36 \pm 0.15$\,mag), based on the tip of the red giant branch (TRGB) method \citep{Monachesi2016MNRAS.457.1419M}.
The Galactic reddening toward AT\,2021biy is very small, $E(B-V)_{\rm MW} = 0.015$\,mag \citep{Schlafly2011ApJ...737..103S}. To estimate the host-galaxy extinction, we measured the equivalent width (EW) of the narrow interstellar Na\,{\sc i}\,D $\lambda\lambda$5890,5896 absorption at the redshift of NGC\,4631  in our earliest spectrum with good signal-to-noise ratio (S/N; see Sect. \ref{specevol})\footnote{We remark that the EW of Na\,{\sc i}\,D was measured in the earliest spectrum, which is not contaminated by spectral features intrinsic to the LRN ejecta. The apparent change with time of that feature is discussed in Sect. \ref{specevol}. }, and found it to be EW $=1.6\pm0.6$\,\AA.
\citet{Poznanski2012MNRAS.426.1465P} provides an empirical relation between Na\,{\sc i}\,D EW and dust extinction, however, it saturates at EW beyond 0.8 \,\AA. Therefore, following \citet{Turatto2003fthp.conf..200T}, we infer a host-galaxy reddening of $E(B-V)_{\rm host}=0.256\pm0.096$\,mag; thus, the total colour excess is $E(B-V)_{\rm total} = 0.271 \pm 0.096$\,mag toward AT\,2021biy. As a consistency check, we measured the Balmer decrement in the earliest spectra of AT\,2021biy after correction for the above reddening amount, and found a value for the \Ha/\Hb\ ratio consistent with the Case~B recombination value of 2.86 \citep[e.g.][]{Weedman1977ARA&A..15...69W}.

\section{Photometry} \label{sec:photometry}

\subsection{Facilities and data reduction} \label{subsec:datareduction}
Follow-up images in the Johnson-Cousins $UBV$, Sloan $ugriz$, and NIR $JHK$ filters were obtained using a number of facilities available to our collaboration. Their setups are summarised as follows:
The Las Cumbres Observatory (LCO; \citealt{Brown2013PASP..125.1031B}) global telescopes located at different sites\footnote{These photometric data come from the Global Supernova Project.}: firstly, ELP (two 1\,m telescopes) at McDonald Observatory, Texas, USA. secondly, TFN (two 0.4\,m telescopes and one 1\,m telescope) at Teide Observatory, Tenerife, Spain;
 The 0.8\,m Tsinghua-NAOC Telescope (TNT) at Xinglong Observatory, China;
The 0.67\,m/0.92\,m Schmidt telescope with a Moravian camera at Padova Astronomical Observatory, Istituto Nazionale di Astrofisica (INAF), Asiago, Italy;
 The 1\,m Zeiss telescope of the Special Astrophysical Observatory (SAO), Russian Academy of Sciences (RAS), Russia;
The 1.82\,m Copernico Telescope with the Asiago Faint Object Spectrograph and Camera (AFOSC), hosted by INAF -- Padova Astronomical Observatory, at the Asiago site, Italy; 
 The 2.0\,m Liverpool telescope (LT) equipped with the IO:O camera, located at Observatorio Roque de Los Muchachos, La Palma, Spain;
The 2.56\,m Nordic Optical Telescope (NOT), at Observatorio Roque de Los Muchachos, La Palma, Spain, with the Alhambra Faint Object Spectrograph and Camera (ALFOSC) and the Nordic Optical Telescope near-infrared Camera (NOTCam);
 The 2.5\,m Caucasus telescope with the IR camera ASTRONIRCAM \citep[][]{Nadjip2017AstBu..72..349N}, hosted by the Caucasian Observatory of the Sternberg Astronomical Institute (SAI) of Lomonosov Moscow State University; 
The 3.5\,m telescope with the Omega-2000 NIR imager at the Calar Alto Observatory, Spain;
 The 6\,m telescope (BTA -- Big Telescope Alt-azimuth) equipped with the SCORPIO-1 and SCORPIO-2 instruments, located near Mt. Pastukhova of the Special Astrophysical Observatory, Russia;
The 10.4\,m Gran Telescopio Canarias (GTC), at Observatorio Roque de Los Muchachos, La Palma, Spain, with the Espectr\'{o}grafo Multiobjeto Infra-Rojo (EMIR) instrument;
 One epoch of mid-infrared (MIR) photometry in the $W1$ and $W2$ filters was obtained by the Wide-field Infrared Survey Explorer (\textit{WISE}) spacecraft.

Optical raw images were first corrected for bias, overscan, and flat-field, following standard steps in the \textsc{iraf}\footnote{\url{https://iraf-community.github.io/}} environment \citep{Tody1986SPIE..627..733T, Tody1993ASPC...52..173T}. If multiple exposures were observed in one night, they were median combined to increase the S/N. The photometric measurements were performed using a \textsc{Python}-based pipeline, {\sl SNOoPY}\footnote{{\sl SNOoPY} is a package developed by E. Cappellaro for performing SN photometry using point-spread-function (PSF) fitting and/or template subtraction. A package description can be found at \url{http://sngroup.oapd.inaf.it/snoopy.html}}, which consists of several packages for photometry, such as {\sc sextractor}\footnote{\url{www.astromatic.net/software/sextractor/}} \citep{Bertin1996A&AS..117..393B}, {\sc daophot}\footnote{\url{http://www.star.bris.ac.uk/~mbt/daophot/}} \citep{Stetson1987PASP...99..191S}, and {\sc hotpants}\footnote{\url{http://www.astro.washington.edu/users/becker/v2.0/hotpants.html/}} \citep{Becker2015ascl.soft04004B}. 
Given that AT\,2021biy was strongly contaminated by the host-galaxy background, the template-subtraction technique was applied in our measurements. These templates were obtained through public archives, such as the Sloan Digital Sky Survey \citep[SDSS; $u$ image taken in 2004;][]{Adelman-McCarthy2006ApJS..162...38A} and the Panoramic Survey Telescope and Rapid Response System \citep[Pan-STARRS; $griz$ images in 2012;][]{Chambers2016arXiv161205560C} for Sloan filters, and the Isaac Newton Telescope (INT; $UBV$ images in 2003) for Johnson-Cousins filters. 

Specifically, the instrumental magnitudes were measured through the PSF-fitting technique. A PSF model was built by fitting the profiles of bright, isolated stars in the field of the transient. Then, the fitted source was removed from the original images, and the residuals at the object's location were used to evaluate the quality of the fits. If the target source was not detected, a magnitude limit was estimated. 

We used zero points (ZPs) and colour terms (CTs) of individual instruments to calibrate the instrumental magnitudes, which were determined through observations of standard photometric fields during photometric nights. Specifically, Johnson-Cousins-filter images were calibrated via the \citet{Landolt1992AJ....104..340L} catalogue, while Sloan-filter data were retrieved from the SDSS DR~13 catalogue \citep{Albareti2017ApJS..233...25A}. In order to improve the calibration accuracy, a local sequence of stars in the transient's field was used to correct the instrumental ZPs during nonphotometric nights. Their catalogued Sloan magnitudes were directly taken from SDSS, while the Johnson-Cousins magnitudes of the reference stars were derived from the Sloan magnitudes after applying the conversion relations of \citet{Chonis2008AJ....135..264C}. 

Photometric errors were estimated through artificial-star simulations, in which several fake stars with known magnitudes (similar to that of the transient) were placed near the position of AT\,2021biy. The magnitudes of the simulated stars were also measured with the PSF procedure. The standard deviation of the magnitudes of the artificial-star experiment was combined (in quadrature) with the PSF-fit and ZP calibration errors, hence providing us with the total photometric uncertainties. The resulting optical magnitudes are reported in electronic form at the CDS.  

NIR data were reduced following similar prescriptions as the optical ones. Raw images were prereduced with flat fielding, distortion correction, and sky subtraction, and then combined to increase the S/N. Instrumental magnitudes were measured via the PSF-fitting technique, and finally, the apparent magnitudes were calibrated using the catalogue of the Two Micron All Sky Survey \citep[2MASS\footnote{\url{http://irsa.ipac.caltech.edu/Missions/2mass.html/}};][]{Skrutskie2006AJ....131.1163S}. The resulting NIR magnitudes are given at the CDS.

\subsection{Multiband light curves of AT\,2021biy} \label{subsec:applc}

\begin{figure*}[htb]
\centering
\includegraphics[width=.95\textwidth]{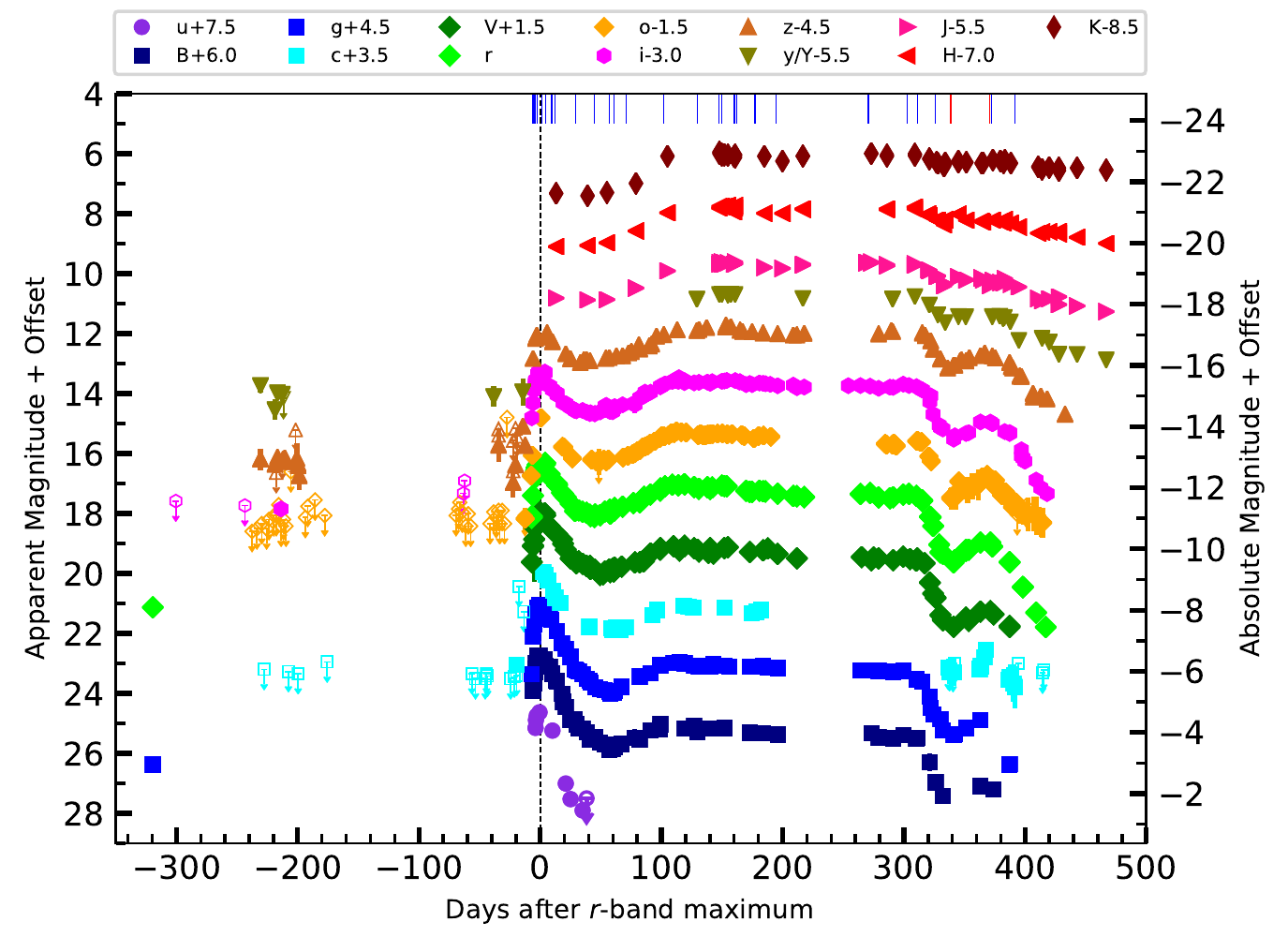}
\caption{Multiband light curves of AT\,2021biy. The dashed vertical line indicates the $r$-band maximum light as the reference epoch. The epochs of our optical spectra are marked with vertical blue solid lines, while the two epochs of NIR spectra are indicated with vertical red solid lines. Upper limits are marked by empty symbols with downward-pointing arrows. For clarity, individual-band light curves are shifted with arbitrary constants reported in the legend. In most cases, the data-point uncertainties are smaller than the plotted symbols. }
\label{applc}
\end{figure*}

Although AT\,2021biy was discovered by ATLAS on MJD = 59243.56 ($\sim 7$\,days before maximum brightness), we searched for earlier survey data from the forced-photometry servers of ATLAS \citep{Smith2020PASP..132h5002S} and Pan-STARRS \citep[][]{Chambers2016arXiv161205560C, Magnier2020ApJS..251....6M}. This allows us to constrain the prediscovery evolution of AT\,2021biy. ATLAS provides magnitudes in the cyan and orange filters, while Pan-STARRS gives $iz$ (very close to Sloan $iz$ filters) and $y$-band data. In this paper, the above magnitudes are maintained in their original photometric systems. The ATLAS and Pan-STARRS magnitudes are made available at the CDS.

ATLAS monitored the site of AT\,2021biy on MJD = 57364.61 ($\sim 1886$\, days prior to maximum brightness), providing a first upper limit of $o>19.2$\,mag. Pan-STARRS DR1 (PS1) reported an observation ($y>19.7$\,mag) on MJD = 56475.26 ($\sim 2776$\,days before maximum). By inspecting these prediscovery data, no detections are reported above the 3$\sigma$ threshold from about 2776 to 231\,days before maximum. The earliest detection of AT\,2021biy in PS1 is in fact on MJD = 59020.26 ($\sim 231$\,days before maximum) at $z=20.70\pm0.35$\,mag (Fig.~\ref{applc}). An earlier detection (MJD = 58931.46; about $-320$\,days) is recovered in archival images from the 3.6\,m Canada-France-Hawaii Telescope (CFHT; equipped with MegaPrime), with $g = 21.88 \pm 0.08$\,mag and $r = 21.13 \pm 0.11$\,mag. 

Afterward, several additional detections were registered during the pre-outburst phase (Fig.~\ref{applc}); there is little, if any, evidence of variability, in contrast with what was found for the Galactic LRN~V1309\,Sco\footnote{V1309\,Sco exhibits a slow rise in luminosity a few months prior to its outburst, consistent with the photometric evolution of the system after the common envelope ejection.} \citep{Tylenda2011A&A...528A.114T}.

Soon after the discovery, we started a high-cadence follow-up campaign lasting over 400\,days. In addition, the subsequent outburst of AT\,2021biy was well monitored by ATLAS and PS1. The multiband light curves of AT\,2021biy, including  pre-outburst data extending up to $\sim 1$\,yr before the LRN discovery, are shown in Figure~\ref{applc}. 
Similar to other LRNe, AT\,2021biy rapidly rises to the light-curve peak in all bands in $\sim 7$\,days. We performed a fifth-order polynomial fit to the early-time $r$-band light curve 
to derive the time of $r$ maximum on MJD = $59251.0 \pm 1.0$ ($r_{\rm{max}} = 16.28 \pm 0.03$\,mag), which will be used throughout this paper. The peak is followed by a rapid decline (with a rate of $4.58 \pm 0.29$\,mag\,$(100\,{\rm d})^{-1}$ in $r$) lasting about 50\,days and reaching a shallow minimum at $r_{\rm{min}} = 18.04 \pm 0.01$\,mag. A plateau is then observed with an average magnitude of $r_{\rm{plateau}} = 17.19 \pm 0.14$\,mag and a duration of 210\,days. Note that a plateau or a broad second peak are typical features of LRNe \citep[see, e.g. the sample of][]{Pastorello2019A&A...630A..75P}. After the plateau, AT\,2021biy shows a rapid drop in all optical bands (of $\sim 2$\,mag in $B$), followed by a short-duration bump in optical light curves. The NIR light curves, although lacking early-time observations, appear to evolve similar to the optical light curves, with a plateau duration of $\sim 210$\,days. After a modest decline, $\sim 310$\,days from $r_{\rm{max}}$ the NIR light curves have another pseudoplateau lasting until our latest observations at $\sim 470$\,days. The slopes, fitted with a linear function, at different sections of the light curve are reported in Table \ref{lc_para}.

\begin{figure}[htb]
\includegraphics[width=0.48\textwidth]{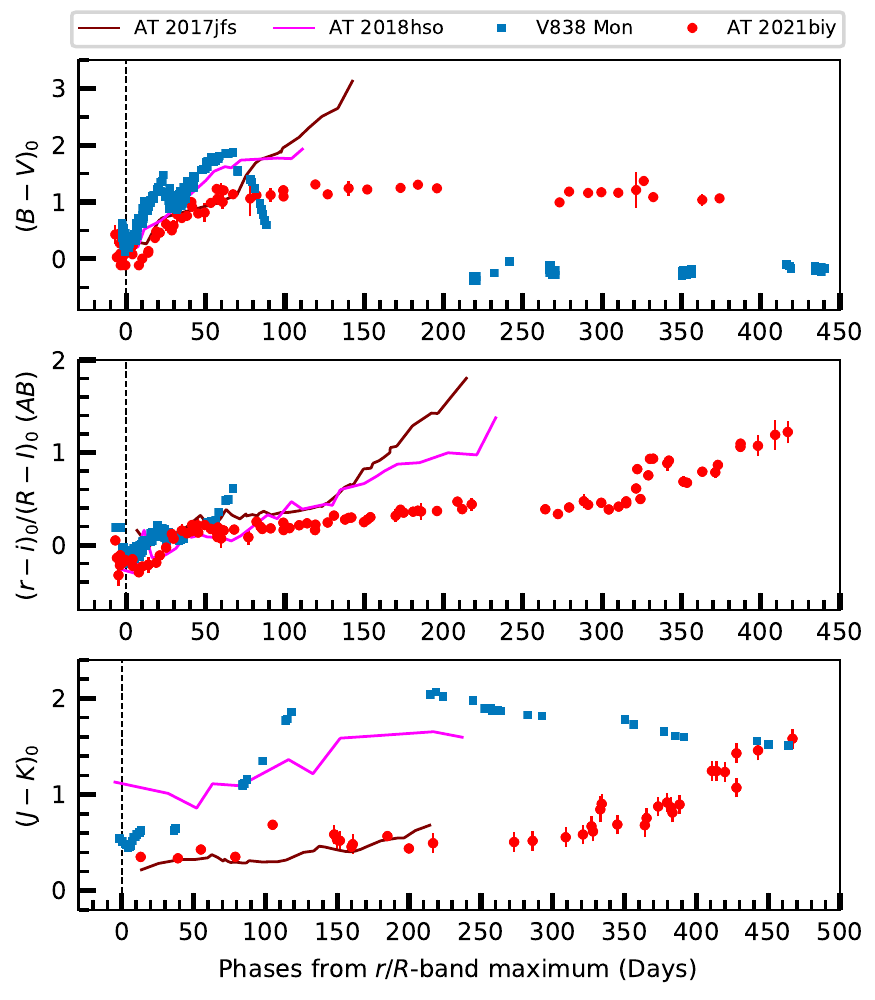}
\caption{Colour evolution of AT\,2021biy compared with that of LRNe AT\,2017jfs, AT\,2018hso, and V838\,Mon. The colours are corrected for extinction, adopting the values reported by the referenced publications. }
\label{colorlc}
\end{figure}

\subsection{Light-curve comparison with other LRNe}

\begin{figure*}[htb]
\centering
\includegraphics[width=0.9\textwidth]{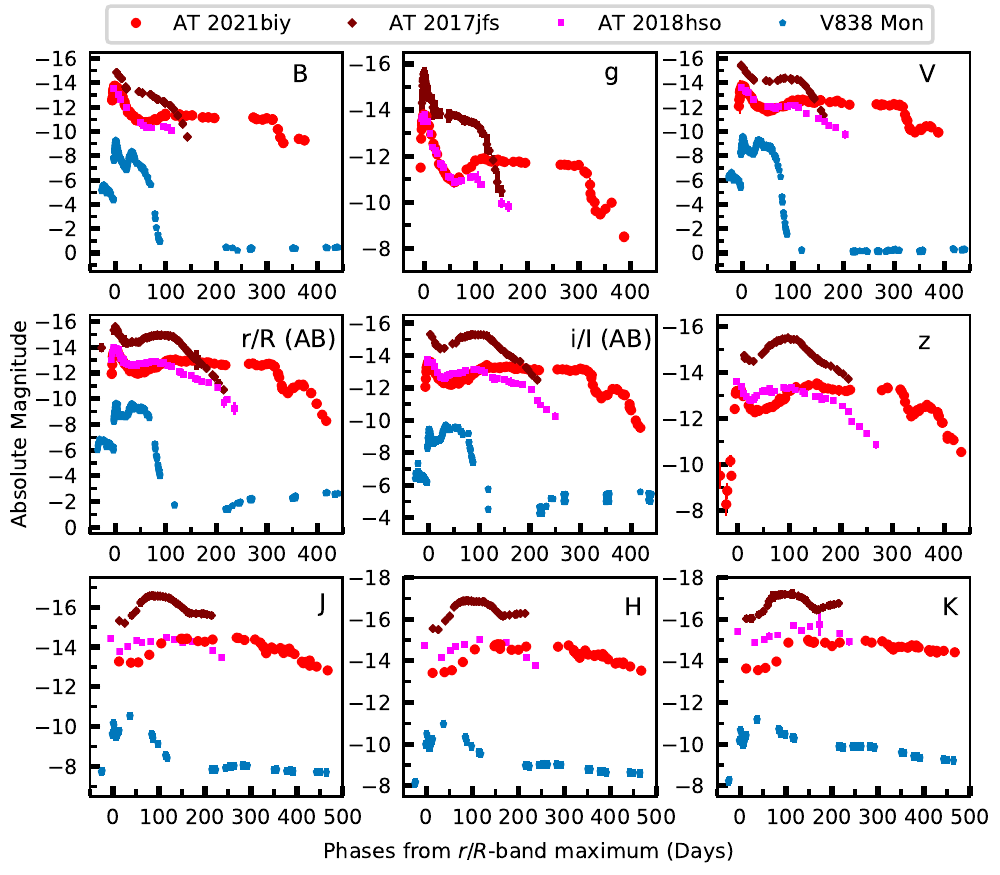}
\caption{Multiband ($B,g,V,r/R,i/I,z,J,H,K$) absolute light curves of AT\,2021biy, compared with those of LRNe AT\,2017jfs, AT\,2018hso, and V838\,Mon.}
\label{abslc}
\end{figure*}

\begin{figure*}[htb]
\centering
\includegraphics[width=.48\textwidth]{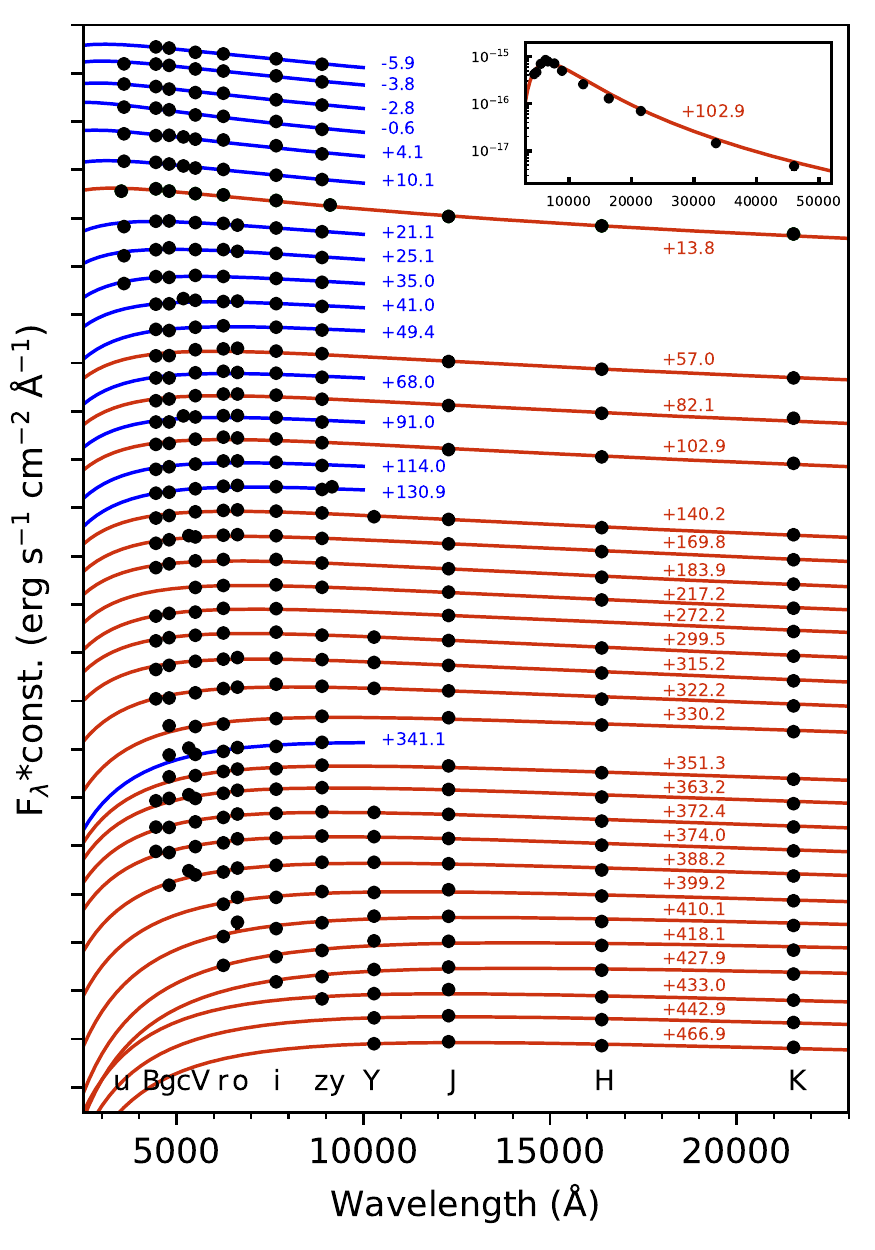}
\includegraphics[width=.48\textwidth]{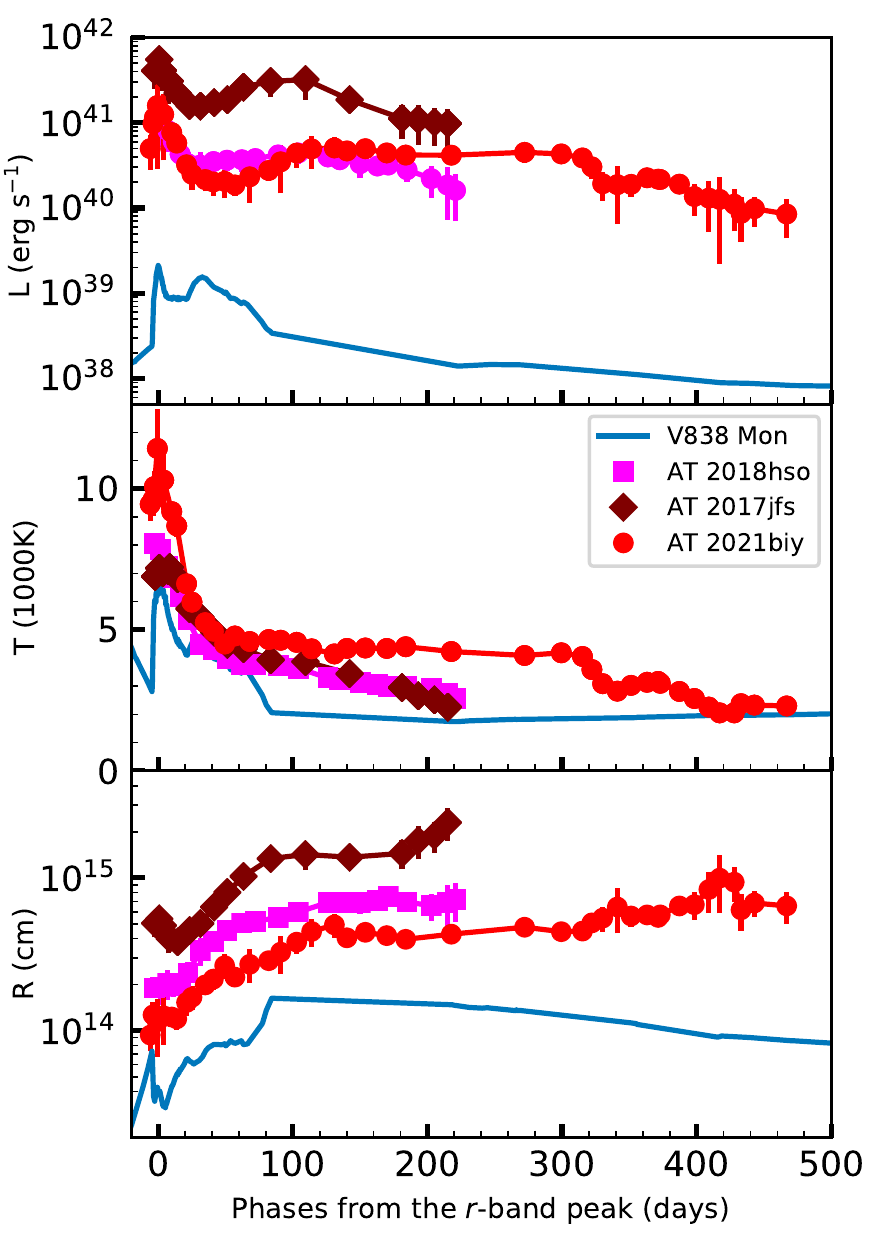}
\caption{Black-body fits for AT\,2021biy with the multi-band observed data. {\sl Left}: SED evolution of AT\,2021biy; for each SED, the best-fit single BB function is also shown. The +102.9\,d SED extended to the MIR thanks to the {\it WISE} data is shown in the inset. Blue lines represent the fits built with the optical data, while red lines are the fits based on the optical and NIR data. {\sl Right}: BB parameters of AT\,2021biy, AT\,2017jfs, AT\,2018hso, and V838\,Mon. {\sl Top-right}: bolometric light curve; {\sl Middle-right}: BB temperature evolution; {\sl Bottom-right}: BB radius evolution.}
\label{pic:SEDs}
\end{figure*}

The colour evolution of AT\,2021biy, along with that of comparison LRNe including AT\,2017jfs \citep[][]{Pastorello2019A&A...625L...8P}, AT\,2018hso \citep[][]{Cai2019A&A...632L...6C}, and the Galactic V838\,Mon event \citep[][]{Goranskii2002AstL...28..691G, Munari2002A&A...389L..51M, Tylenda2005A&A...436.1009T}, is shown in Figure~\ref{colorlc}. At the earliest epochs (about $-7$\,days), the intrinsic $(B-V)_0$ and $(r-i)_0$ colours of AT\,2021biy are $\sim 0.5$ and $\sim 0.1$\,mag, respectively. Then, the object evolves to bluer colours, with $(B-V)_0 \approx -0.2$\,mag and $(r-i)_0 \approx -0.4$\,mag at about $+10$\,days. These colour minima are followed by a new rise from $+$10 to $+$50\,days, and then by a flattening with a duration of $\sim 260$\,days, reaching colours of $(B-V)_0 \approx 1.0$\,mag and $(r-i)_0 \approx 0.5$\,mag at about $+$310\,days. This slow colour evolution is also consistent with that of the temperature (see Sect. \ref{subsec:sed}). Accompanied by the late-time light-curve rebrightening, the colours of AT\,2021biy show a short-duration ($\sim 1$\,month) red bump starting at $\sim 330$\,days. After the rebrightening phase, $(r-i)_0$ rapidly reddens again, rising up to $\sim 1.2$\,mag at $\sim 420$\,days. 
The $(J-K)_0$ colour evolution is similar to that of $(B-V)_0$ and $(r-i)_0$, getting steadily redder from 0.3\,mag to 1.6\,mag up to 470\,days. As shown in Figure~\ref{colorlc}, the colour evolution of AT\,2021biy resembles that of the comparison LRNe but with different timescales. A diverse flattening duration in the colour curves can be comfortably explained with different timescales in the hydrogen recombination front moving across the expanding gas \citep[e.g.][]{Ivanova2013A&ARv..21...59I,Lipunov2017MNRAS.470.2339L}.

In Figure~\ref{abslc}, we show the absolute light curves of AT\,2021biy and the comparison LRNe. All show a similar morphology, with the double-hump light curves being the most characteristic LRN feature. However, a diversity of luminosity peaks (both first and second peaks during the outburst phase) and plateau durations is observed in LRNe \citep[see a sample of LRNe in ][]{Pastorello2019A&A...630A..75P, Pastorello2021A&A...646A.119P, Stritzinger2020A&A...639A.104S}. 
During the major outburst, AT\,2021biy reaches $M_r = -13.92 \pm 0.23$\,mag, comparable to $M_r \approx -13.93$\,mag for AT\,2018hso, while it is fainter than AT\,2017jfs with $M_r \approx -15.74$\,mag. We note that AT\,2021biy shows a long-lasting plateau of $\sim 210$\,days (with $M_r = -13.12 \pm 0.23$\, mag) resembling that of Type IIP SNe \citep[][]{Popov1993ApJ...414..712P}, rather than a broad, red second peak observed as in most LRNe of our sample (see Fig.~\ref{abslc}). This variety is likely due to the different masses of the involved ejecta and/or surrounding circumstellar envelope \citep[e.g.][]{Stritzinger2020A&A...639A.104S,Pastorello2021A&A...646A.119P,Pastorello2021A&A...647A..93P}.

After the plateau, AT\,2021biy displays a rapid luminosity drop in the optical domain (e.g. $\gamma_4 (r) = 7.94 \pm 0.70$\,mag\,$(100\,{\rm d})^{-1}$, while a significantly slower decline is observed in the NIR bands (e.g. $\gamma_4 (H) = 2.08 \pm 0.32$\,mag\,$(100\,{\rm d})^{-1}$). Remarkably, a late rebrightening of AT\,2021biy is visible both in the optical and NIR domains, analogous to that observed in V838\,Mon (Fig.~\ref{abslc}). However, AT\,2021biy shows a short-duration hump in the light curve, much shorter than that observed in V838\,Mon. Late-time fluctuations in the light curve of V838\,Mon were tentatively associated with the effect of the fragmentation of the dusty envelope and the emerging light of the merger outcome \citep[][]{Munari2002A&A...389L..51M}. This explanation appears to be less plausible for AT\,2021biy, where a short-lived light-curve hump can also be produced by a collision of ejected material with a relatively thin gas shell.

\subsection{Bolometric light curve, temperature, and radius} \label{subsec:sed}
The evolution of the spectral energy distribution (SED) of AT\,2021biy is constructed from the observed light curves, adopting the distance and extinction values reported in Sect. \ref{sec:info}, and following the procedures implemented for other published LRNe \citep[e.g. AT\,2018hso;][]{Cai2019A&A...632L...6C}. In order to study the evolution of the bolometric luminosity, we fitted the SEDs at some representative epochs with single black-body (BB) functions (see the left panel of Fig.~\ref{pic:SEDs}). The bolometric luminosity for each SED is computed by integrating the BB flux over the entire electromagnetic spectrum. The resulting bolometric light curve, along with the inferred evolution of the BB temperature and radius, is shown in the right panel of Figure~\ref{pic:SEDs}, while their values are reported in Table \ref{AT2021biySED}.

At the first blue peak, AT\,2021biy has a bolometric luminosity of $\sim 1.6 \times 10^{41}$\,erg\,s$^{-1}$, which is a factor of 3.4 fainter than AT\,2017jfs ($\sim 5.5 \times 10^{41}$\,erg\,s$^{-1}$) but slightly brighter than AT\,2018hso ($\sim 1.1 \times 10^{41}$\,erg\,s$^{-1}$; see the top-right panel of Fig.~\ref{pic:SEDs}). After a fast post-peak decline with a minimum luminosity of $\sim 1.9 \times 10^{40}$\,erg\,s$^{-1}$, AT\,2021biy rises to a plateau at $\sim 5.0 \times 10^{40}$\,erg\,s$^{-1}$, while AT\,2017jfs and AT\,2018hso reach their second red maximum at $\sim 3.2 \times 10^{41}$\,erg\,s$^{-1}$ and $\sim 4.4 \times 10^{40}$\,erg\,s$^{-1}$, respectively. After the plateau (or the second peak), the luminosities of all objects decline again. Later, AT\,2021biy exhibits a short-duration hump in the bolometric light curve, also observed in V838\,Mon.

The BB temperature of AT\,2021biy rapidly rises to a peak of 11,450\,K (middle-right panel of Fig.~\ref{pic:SEDs}), but soon after maximum brightness, it drops very quickly to 4500\,K at +50\,days. Later, the temperature evolves slowly, cooling to 4050\,K at $+$315\,days, supporting the H recombination explanation for the light-curve plateau. The comparison LRNe show a similar evolution trend with AT\,2021biy during this stage; however, AT\,2017jfs ($T_{\rm{BB}} \approx 7000$\,K) and AT\,2018hso ($T_{\rm{BB}} \approx 8000$\,K) are much cooler at their maxima. After the plateau, the temperature of AT\,2021biy initially declines to a local minimum of $T_{\rm{BB}} \approx 2850$\,K and then rises again to $T_{\rm{BB}} \approx 3150$\,K at +371\,days. Finally, it declines again to $T_{\rm{BB}} \approx 2300$\,K at our last observations around +467\,days. 
In contrast, the Galactic LRN V838\,Mon shows a very slow temperature rise at very late phases, going from $\sim 1750$\,K to $\sim 2000$\,K.

Figure~\ref{pic:SEDs} (bottom-right panel) shows the radius evolution of the four LRNe, which are inferred through the Stefan-Boltzmann law ($L=4\pi R^2 \sigma T^4$, where $\sigma$ is the Stefan-Boltzmann constant) using the luminosity and temperature parameters estimated above. LRNe show relatively homogeneous radius evolution before the rebrightening phase, which is suggested to be a diagnostic tool to discriminate LRNe from the so-called ILRTs \citep[see the discussion by][]{Cai2019A&A...632L...6C}. 
At maximum light, AT\,2021biy has the smallest photospheric radius ($1.1 \times 10^{14}$\,cm = 1580\,\rsun) among the comparison LRNe: AT\,2017jfs ($5.4 \times 10^{14}$\,cm = 7760\,\rsun) and AT\,2018hso ($1.9 \times 10^{14}$\,cm = 2730\,\rsun). After maximum brightness, the photospheric radius of AT\,2021biy initially remains roughly constant, then steadily rises to $4.9 \times 10^{14}$\,cm (7040\,\rsun) at +131\,days; it expands at $\sim 335$\,\kms. This is followed by a relatively slow evolution, with $R \approx 4.5 \times 10^{14}$\,cm (6470\,\rsun) until $+$315\,days. 
The radius then reaches a local maximum of $6.4 \times 10^{14}$\,cm (9200\,\rsun) at +341\,days and evolves to $R \approx 6.5 \times 10^{14}$\,cm (9340\,\rsun) at the latest monitored epoch. In this phase, the SED peak appears to be shifted from the optical to the IR domains. According to current observations, AT\,2021biy is showing an opposite trend in comparison with the very late-time radius evolution of V838\,Mon, as the latter had a slow radius decline up to $R \approx 8.6 \times 10^{13}$\,cm (1240\,\rsun) at $+$470\,days.

\section{Spectroscopy}  \label{sec:spectra}  

\begin{figure*}[htbp]
\centering
\includegraphics[height=235mm,width=145mm]{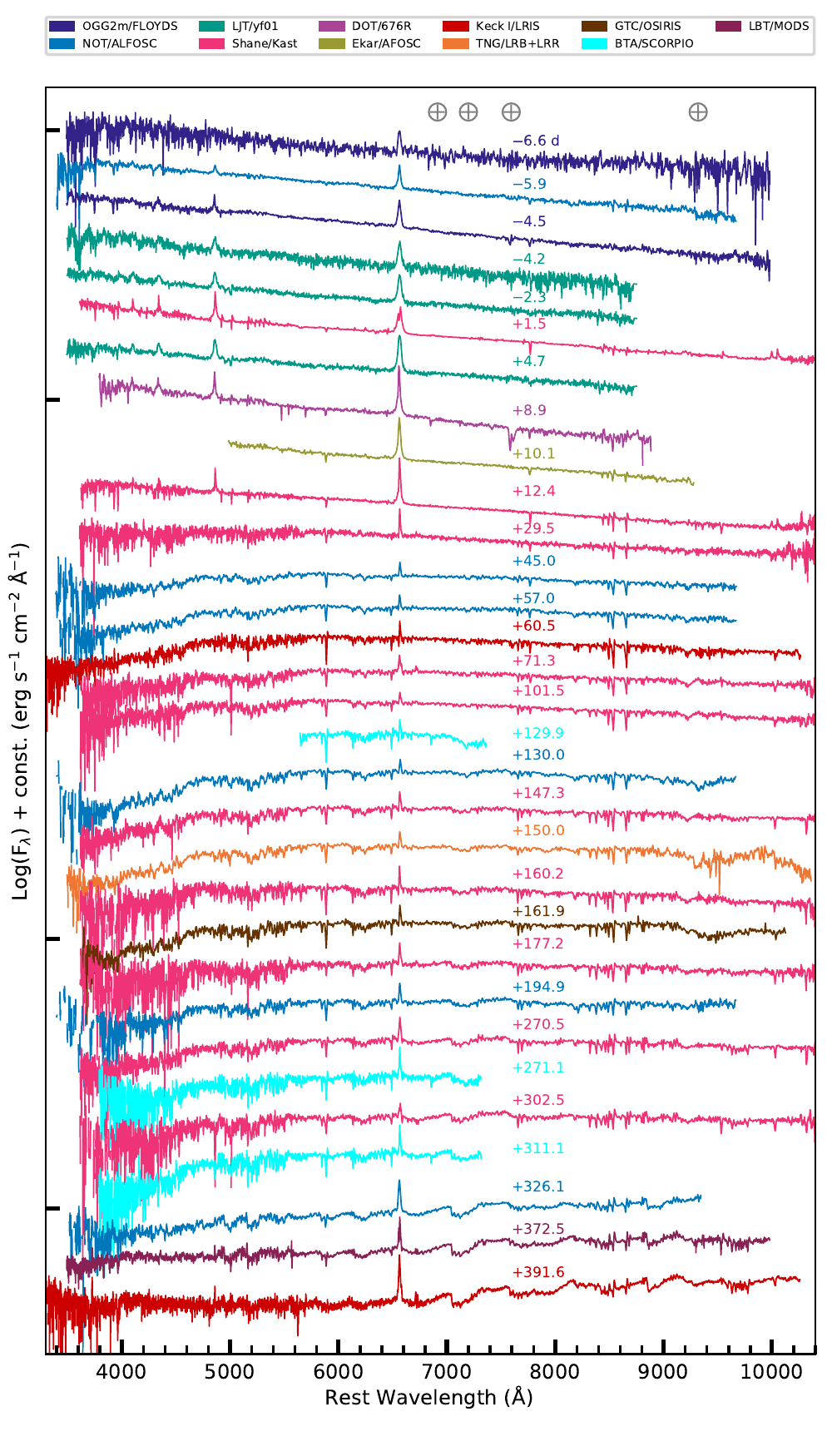}
\caption{Spectral evolution of AT\,2021biy. The spectra have been corrected for redshift and reddening. The symbol $\bigoplus$ marks the locations of strong telluric absorption bands. The phases extend from $-6.6$ to $+$391.6\,days.}
\label{spectseq}
\end{figure*}

\begin{figure}[htb]
\includegraphics[width=0.5\textwidth]{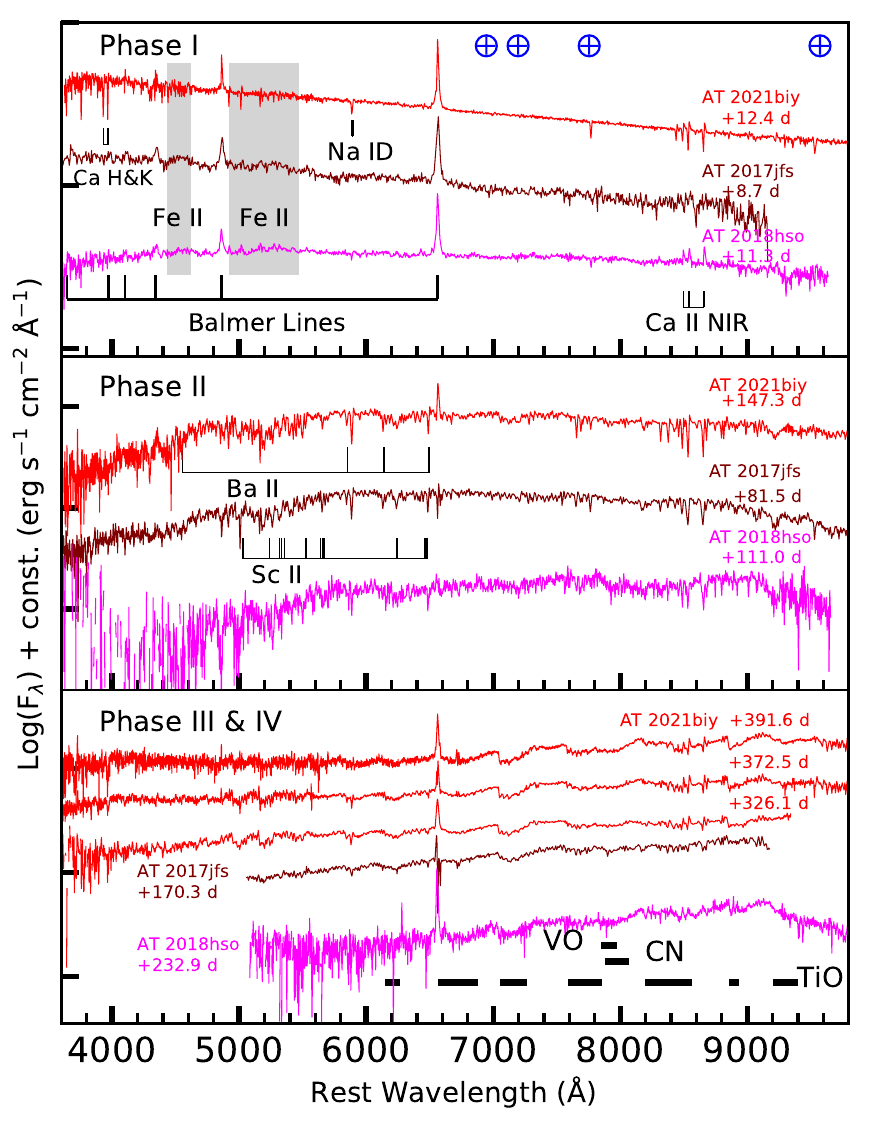}
\caption{Spectral comparison of AT\,2021biy with LRNe AT\,2017jfs and AT\,2018hso. The spectra were selected at three distinctive phases: {\sl Phase} {\sc i} -- soon after the first blue peak ({\sl top panel}); {\sl Phase} {\sc ii} -- around the second red peak/plateau ({\sl middle panel}); and {\sl Phase} {\sc iii} -- at late epochs ({\sl bottom panel}). Two spectra taken during the late light-curve hump ({\sl Phase} {\sc iv}) are also shown in the {\sl bottom panel}. The principal lines in the LRN spectra are marked, following the identifications of \citet{Pastorello2019A&A...625L...8P}, \citet{Pastorello2019A&A...630A..75P}, \citet{Pastorello2021A&A...646A.119P}, \citet{Pastorello2021A&A...647A..93P}, and \citet{Stritzinger2020A&A...639A.104S}. 
All spectra have been corrected for redshift and reddening.}
\label{SpectraComp}
\end{figure}

\begin{figure}[htb]
\centering
\includegraphics[width=0.4\textwidth]{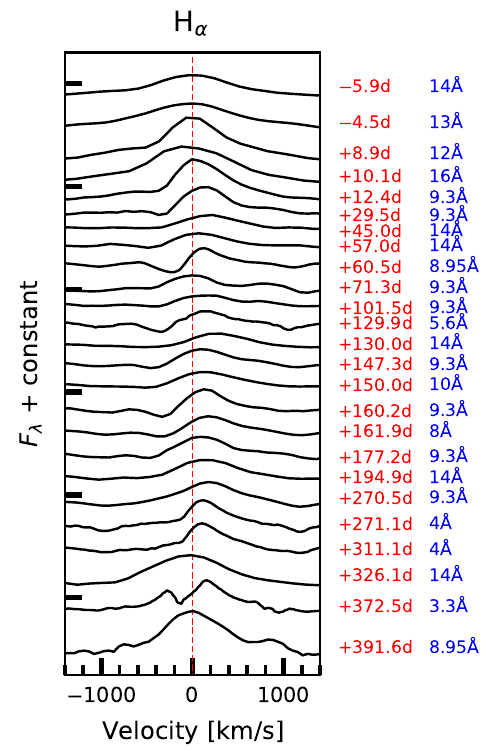}
\caption{Evolution of the \Ha\ profile of AT\,2021biy. The \Ha\ lines are shown in velocity space, with zero velocity marked by a red dotted line. Phases (red colour) and resolution (blue colour) of each spectrum are labeled on the right-hand side. Phases are relative to $r$-band maximum. All of the spectra are corrected for redshift and reddening. Note that some low-S/N spectra are not shown in this figure.}
\label{Hprofiles}
\end{figure}

\begin{figure}[htbp]
\centering
\includegraphics[width=9.0cm]{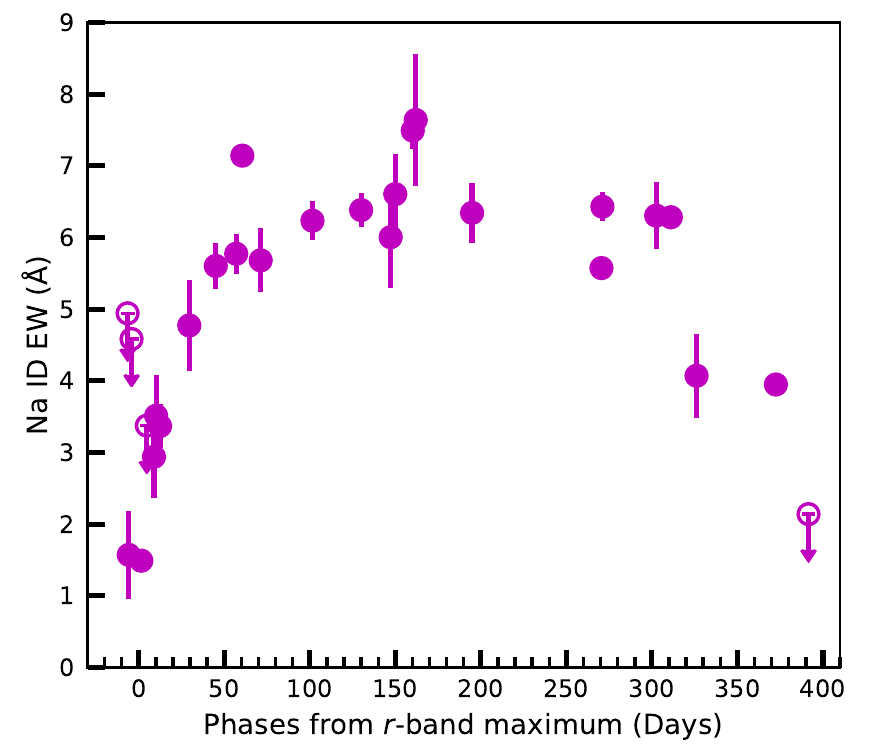}
\caption{Temporal evolution of the Na\,{\sc i}\,D $\lambda\lambda$5890, 5896 EW in the spectra of AT\,2021biy. The Na\,{\sc i}\,D measurements are performed through the Markov Chain Monte Carlo (MCMC) methods using the \textsc{emcee} \textsc{Python} package \citep{ForemanMackey2013PASP..125..306F}. Spectra are trimmed to the region surrounding the doublet and are fit with a double Gaussian subtracted from a linear continuum. The MCMC implementation initialises walkers for each parameter of the model and iteratively perturbs each parameter, checking at each step whether the fit to the data has improved. Upper limits are marked by empty symbols with downward-pointing arrows.}
\label{NaID_EW}
\end{figure}

\begin{figure*}[htb]
\centering
\includegraphics[width=0.96\textwidth]{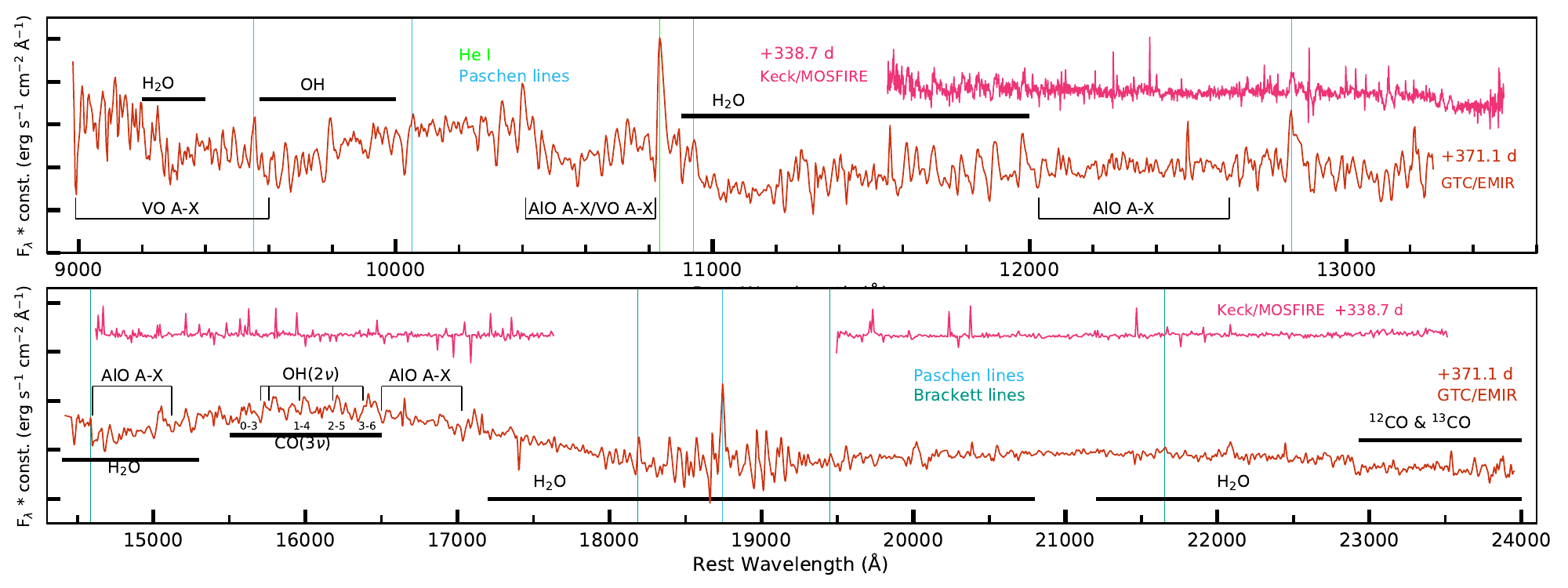}
\caption{Two NIR spectra of AT\,2021biy taken at $+$338.7\,days and $+$371.1\,days. The $H$- and $K$-band spectra taken with Keck-I/MOSFIRE have been binned to 8\,\AA\ and 9\,\AA\ respectively owing to their extremely low S/N.}
\label{nirspec}
\end{figure*}

\begin{figure}[ht]
\centering
\includegraphics[width=0.52\textwidth]{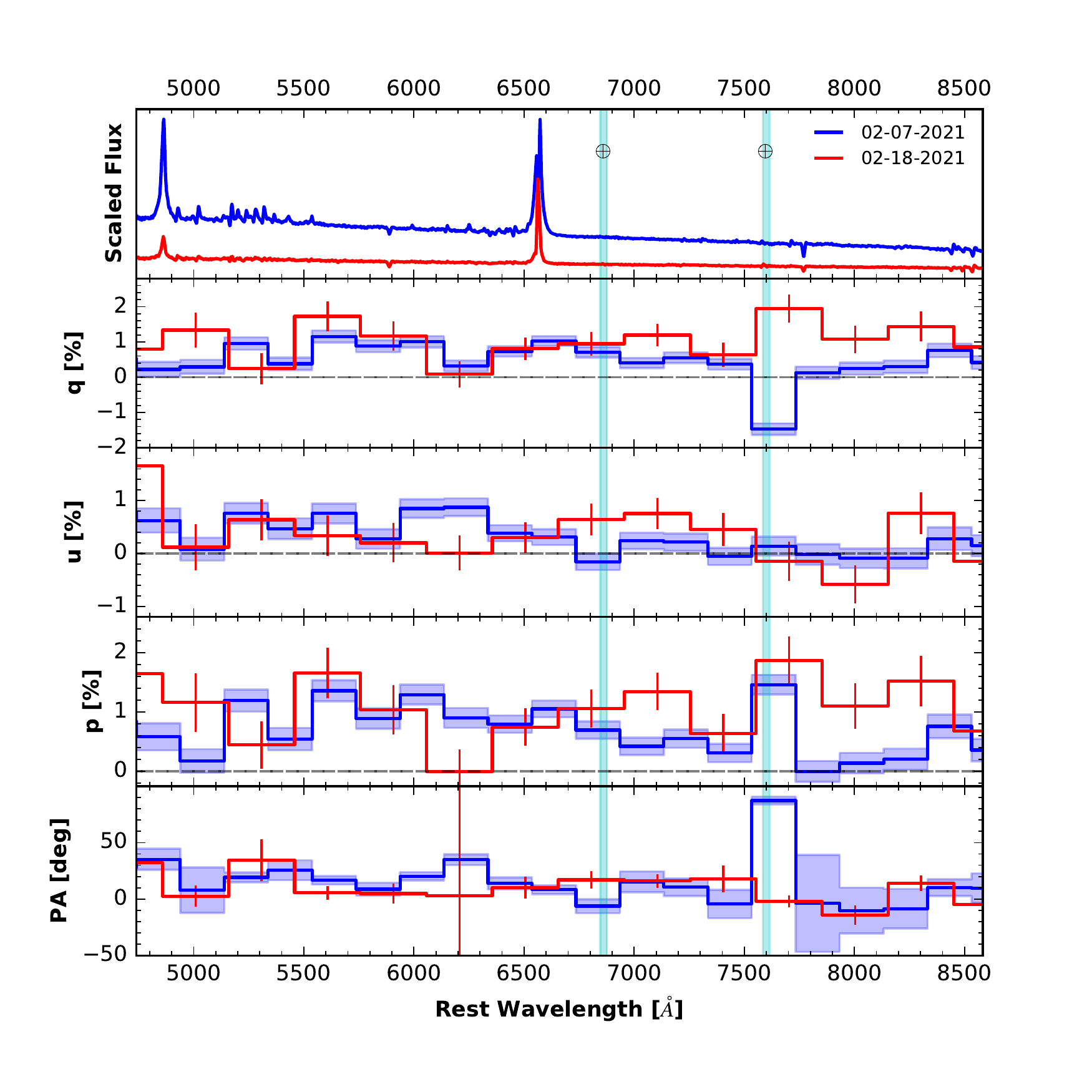}
\caption{Spectropolarimetry of AT\,2021biy. The top panel shows the total-flux spectra in flux arbitrary units. The second, third, and fourth panels display the Stokes $q$, $u$, and polarization $p$ spectra, respectively. The bottom panel gives the polarization position angle, PA. The vertical cyan bands show the regions affected by telluric bands. With the exception of the total-flux spectra, data are binned to 200\,\AA\ and 300\,\AA\ for the first and second epochs (respectively), for clarity of presentation.}
\label{specpol}
\end{figure}

High-cadence spectroscopic observations of AT\,2021biy were performed using multiple instrumental configurations. Information on the obtained spectra  is reported in Table~\ref{2021biyspecinfo}.
The spectra were reduced following standard procedures in {\sc iraf}.
After the traditional prereduction steps, such as bias and overscan corrections, trimming, and flat fielding, we extracted one-dimensional spectra from the two-dimensional frames. Wavelength and flux calibrations were performed using spectra of comparison lamps and spectrophotometric standard stars, respectively. The calibration images were taken during the same night and with the same instrumental configuration as the LRN spectra. The spectral flux was improved by checking the coeval broadband photometry, and the telluric absorption bands (e.g. O$_2$ and H$_2$O) were removed using the spectra of standard stars. Spectra obtained with the 6\,m BTA equipped with the SCORPIO-1 and SCORPIO-2 multimode focal reducers were calibrated following the descriptions by \citet[][]{Afanasiev2005AstL...31..194A} and \citet{Afanasiev2011BaltA..20..363A}.

A series of optical spectra of AT\,2021biy was obtained with the Kast Double Spectrograph \citep{Miller1993} mounted on the 3\,m Shane telescope at Lick Observatory. The spectra were taken at or near the parallactic angle to minimise slit losses caused by atmospheric dispersion \citep{Filippenko1982PASP...94..715F}. Data reduction followed standard techniques for CCD frame processing and spectrum extraction \citep{Silverman2012MNRAS.425.1789S} using {\sc iraf} routines and custom \textsc{Python} and IDL codes\footnote{\url{https://github.com/ishivvers/TheKastShiv}}. In addition, spectra obtained with Ekar 1.82\,m/AFOSC, 3.6\,m DOT/ADFOSC, NOT/ALFOSC, and GTC/OSIRIS were reduced using the dedicated pipeline {\sl Foscgui} developed by E. Cappellaro, while the Keck {\sc i}/LRIS spectra were processed with the pipeline {\sl LPipe} written by \citet[][]{Perley2019PASP..131h4503P}. The OGG2\,m/FLOYDS spectra were taken as part of the Global Supernova Project. The resulting AT\,2021biy spectral series is shown in Figure~\ref{spectseq}.

\subsection{Spectroscopic evolution}\label{specevol}
We collected 31 optical spectra of AT\,2021biy, covering about 13\,months and all crucial phases of its evolution. They exhibit the typical evolution of LRNe \citep[e.g.][]{Pastorello2019A&A...625L...8P,Pastorello2019A&A...630A..75P,Pastorello2021A&A...646A.119P,Pastorello2021A&A...647A..93P}. We identify four distinct phases: {\sl Phase} {\sc i} -- at early times, around the blue peak; {\sl Phase} {\sc ii} -- an intermediate phase during the plateau; {\sl Phase} {\sc iii} -- at late times, during the fast post-plateau decline; and {\sl Phase} {\sc iv} -- during the late-time light-curve hump.

{\sl Phase} {\sc i}: At early epochs (until $+$29.5\,days after the light-curve maximum), spectra of AT\,2021biy show a blue continuum with superimposed prominent Balmer emission lines. Several Fe\,{\sc ii} emission lines are also observed, along with Ca\,{\sc ii}\,H\&K in absorption and the barely visible Ca\,{\sc ii} NIR triplet (see the top panel of Fig.~\ref{SpectraComp}). The BB temperature, inferred from the spectral continuum, decreases from $\sim 10,800$\,K (at $+$1.5\,days) to $\sim 5850$\,K (at +29.5\,days).

In the NOT/ALFOSC spectrum at $-$5.9\,days, the \Ha\ profile is dominated by an unresolved narrow component superposed on a broader base (Fig.~\ref{Hprofiles}). We thus measure the full width at half-maximum intensity (FWHM) of the \Ha\ emission line through a double-component fit: a narrow Lorentzian and a broader one. The resulting FWHM velocity of \Ha, accounting for the instrumental resolution, has a narrow component with an upper limit of $\sim 640$\,\kms\ and a broader component of $\sim 1680$\,\kms. Soon after maximum brightness, the broader component disappears and the \Ha\ profile is well-reproduced by a single Lorentzian function fit with $v_{\rm{FWHM}} \approx 430$\,\kms\ (close to the Shane/Kast instrumental resolution).

As mentioned in Sect. \ref{sec:info}, we used the early, good-S/N spectrum at phase $-$5.9\,days to measure the EW of the narrow Na\,{\sc i}\,D feature that we attribute to line-of-sight gas within the host galaxy. While EW $= 1.6 \pm 0.6$\,\AA\ in that spectrum, the EW of Na\,{\sc i}\,D seems to change with time. For this reason, we measure the temporal evolution of the EW of Na\,{\sc i}\,D absorption (Fig.~\ref{NaID_EW}); it clearly increases up to a factor of four during the first $\sim 50$\,days, remaining nearly constant during the following 8\,months.  

While the EW and the overall profile of the interstellar Na\,{\sc i}\,D absorption has been observed to change in other types of stellar transients \citep[see, e.g.][]{Patat2007Sci...317..924P, Wang2019ApJ...882..120W}, to our knowledge it has never been observed in LRNe. While we cannot rule out that this is due to changes in the ionisation stage of the line-of-sight interstellar medium or an increased dust condensation (Byrne et al. in prep.), the apparent evolution of this feature in the spectra of AT\,2021biy may have a very simple explanation: the increasing contamination of the Na\,{\sc i}\,D absorption component of the expanding LRN ejecta. The increasing strength of other metal-line absorption seems to support this scenario.

{\sl Phase} {\sc ii}: At later phases, the spectra experience a remarkable change. The spectral continuum initially becomes rapidly redder, from $T \approx 5850$\,K at +29.5\,days to $T \approx 4800$\,K at +101.5\,days; then the temperature decreases more slowly down to $T \approx 3950$\,K at $+$302.5\,days. During the plateau phase, the Balmer emission lines are less prominent than in the early-time spectra (see Fig.~\ref{Hprofiles}) and are similar to those of LRN AT\,2020kog \citep{Pastorello2021A&A...647A..93P}. We note that in most LRNe, the Balmer emission is barely detectable during the plateau, as one can see in the middle panel of Figure~\ref{SpectraComp}. 

In the medium-resolution spectra taken with Keck-I/LRIS (at +60.5\,days) and BTA/SCORPIO-1 (at +129.4\,days), the \Ha\ profiles show a double P~Cygni feature in absorption, with minima blueshifted by $\sim 350$\,\kms\ and $\sim 500$\,\kms, respectively. In addition, a forest of metal lines, such as Fe\,{\sc ii}, Sc\,{\sc ii}, Ba\,{\sc ii}, and Na\,{\sc i}\,D, are observed during this phase (see the middle panel of Fig.~\ref{SpectraComp}). 

{\sl Phase} {\sc iii}: Spectra taken during the rapid post-plateau decline of the light curve show the classical features of old LRNe. The temperature of the spectral continuum further decreases from $T \approx 3950$\,K (at +302.5\,days) to $T \approx 3350$\,K (at +326.1\,days), evolving to that of a late M-type star. \Ha\ becomes prominent again (see Fig.~\ref{Hprofiles}) with a $v_{\rm{FWHM}}$ limit of 730\,\kms\ in the +326.1\,day spectrum. Typical broad molecular absorption bands (mostly TiO, but also VO and CN) emerge in the late-time spectra, which are marked and identified in the bottom panel of Figure~\ref{SpectraComp}. This late-time metamorphosis is consistent with that observed in other LRNe \citep[e.g.][]{Kaminsk2009ApJS..182...33K, Cai2019A&A...632L...6C, Pastorello2019A&A...625L...8P,Pastorello2019A&A...630A..75P,Pastorello2021A&A...646A.119P,Pastorello2021A&A...647A..93P}.

{\sl Phase} {\sc iv}: Soon after the fast luminosity decline, we obtained two spectra during the late-time bumpy phase with LBT/MODS and Keck-I/LRIS. This is the first opportunity to spectroscopically follow the late bumpy phase for an extragalactic LRN. As seen in the bottom panel of Figure~\ref{SpectraComp}, these two spectra still show molecular features superposed on a very red continuum, with $T \approx 3550$\,K (at +372.5\,days) and $T \approx 2950$\,K (at +391.6\,days). The clear detection of a P~Cygni \Ha\ profile in the medium-resolution LBT spectrum at +372.5\,days indicates a velocity of about 130\,\kms\ for an external shell or even the common envelope (see also in Fig.~\ref{Hprofiles}).
The prominent \Ha\ emission line in the Keck-I/LRIS spectrum (at +391.6\,days) is well reproduced by two Gaussian components, narrow and broad, with $v_{\rm{FWHM}} \approx 250$\,\kms\ and $v_{\rm{FWHM}} \approx 1450$\,\kms\ (after correcting for instrumental resolution), respectively. The inspection of the two spectra taken during the late light-curve hump (at around $+$ 370 days; see Fig.~\ref{spectseq} and bottom panel of Fig.~\ref{SpectraComp}) does not unequivocally reveal new features that can be attributed to shock interaction, such as the narrow high-ionization emission lines observed in some strongly interacting supernovae, or broader boxy features typical of shocked regions \citep{Chevalier1994ApJ...420..268C}. This is probably due to the fact that in LRNe the shocked regions are deeply embedded within the envelope \citep[e.g.][]{Metzger2017MNRAS.471.3200M, Aydi2020NatAs...4..776A}. We note, however, that the broader H$\alpha$ wings (with $v_{\rm{FWHM}} \approx 1450$\,\kms\ ) observed in the +391.6\,days spectrum are a plausible signature that higher velocity, shocked material emerged at very late phases.

We also collected two NIR spectra of AT\,2021biy at very late phases (+338.7\,days and +371.1\,days), one with the Keck-I telescope equipped with MOSFIRE and the other with the GTC plus EMIR. This is the second extragalactic LRN published in the literature that has NIR spectroscopy (the first being AT 2018bwo, \citealt{Blagorodnova2021A&A...653A.134B}).
As shown in Figure~\ref{nirspec}, water (H$_{2}$O) vapour and CO are the strongest absorption features, which are similar to those of V838\,Mon at late times \citep[][]{Lynch2007ASPC..363...39L}. By analogy with the optical spectra, molecular absorption features (e.g. AlO, VO) are also detected in the NIR spectra. However, VO absorption is not securely identified in the NIR spectrum. The CO($3\nu$) and OH($2\nu$) overtone absorption bands are clearly detected at $\sim 1.6\,\mu$m in the $H$-band spectrum (Fig.~\ref{nirspec}), which are attributed to the decrease in opacity of the expanding gas. In addition, we identify H emission lines of the Paschen and Brackett series along with the atomic transitions of He\,{\sc i}.

\subsection{Spectropolarimetry}\label{polarimetry} 
Using the Kast spectrograph on the 3\,m Shane telescope at Lick Observatory, we obtained two nights of spectropolarimetry of AT\,2021biy, +1~d and +12~d. Observations and data reduction were carried out as described by \cite{2022MNRAS.509.4058P}. For instrumental calibration, we observed the low-polarization standard star HD\,57702 to confirm the low instrumental polarization on each night. The average fractional Stokes $q$ and $u$ parameters were found to be $<0.05$\%. We also observed the high-polarization standard stars HD\,43384 and HD\,58624, and constrained the polarization and position angle to within 0.1\% and $3^{\circ}$ of the published values, respectively \citep[e.g. ][]{Mathewson1970MmRAS..74..139M, Hsu1982ApJ...262..732H}. Figure~\ref{specpol} shows both epochs of spectropolarimetry of AT\,2021biy.

The observed continuum polarization of AT\,2021biy is $\sim 1$\%. However, it is unclear how much of it is intrinsic to AT\,2021biy and how much is due to dust in the interstellar medium. The `Serkowski law' puts an upper limit on the percent interstellar polarization (ISP) as $9 \times E(B-V)$, where $E(B-V)$ is given in magnitudes \citep{1975ApJ...196..261S}. 
In the direction of AT\,2021biy, the Milky Way extinction is low ($E(B-V) = 0.015$\,mag; \citealt{Schlafly2011ApJ...737..103S}), constraining Galactic ISP to $<0.14$\%. By observing an intrinsically unpolarized star\footnote{Gaia 1514542396422711424} (probe star) that is located within $1^{\circ}$ of the line of sight to AT\,2021biy and at a distance $\sim 700$\,pc away from Earth (to probe a sufficient column of Galactic interstellar medium), we estimated the Milky Way contribution to the interstellar polarization. The probe star was polarized at $<0.12$\%. This value is consistent with the upper limit derived above using the Serkowski law. Therefore, we conclude that the Milky Way contribution to ISP is low compared to the observed polarization. 

As done previously for the LRN V838\,Mon \citep{2004A&A...414..591D}, we fit the polarization spectrum taken on 2021 February 7 with a Serkowski curve to calculate host-galaxy ISP. We note that, being extragalactic, the spectropolarimetry of AT\,2021biy has much lower S/N compared to the Galactic LRN V838\,Mon. We found a maximum polarization $p_{\rm{max}} = 0.93 \pm 0.13$\% for AT\,2021biy, compared with $p_{\rm{max}} = 2.48 \pm 0.10$\% for V838\,Mon. The wavelength at $p_{\rm{max}}$ was determined to be $\lambda_{\rm{max}} = 5360 \pm 910$\,\AA,  consistent within $1\sigma$ of $\lambda_{\rm{max}} = 5750 \pm 50$\,\AA\ for V838\,Mon. However, we note that the constraint on $\lambda_{\max}$ for AT\,2021biy is weaker owing to the low-S/N data. The relative similarity of the values of $\lambda_{\rm{max}}$ suggests that the dust properties in the vicinity of AT\,2021biy are similar to those of dust in the Milky Way in the direction of V838\,Mon. The polarization and position angle are consistent at both epochs, but as a consequence of the lack of additional epochs of polarimetry, we are unable to confidently calculate the intrinsic polarization of AT\,2021biy.

\section{Progenitor analysis} \label{progenitor}

We examined archival data for NGC\,4631 in order to search for the progenitor of AT\,2021biy. While the \textit{Spitzer Space Telecope} has observed this galaxy on multiple occasions between 2004 and 2019, these data unfortunately do not have sufficiently high spatial resolution to search for a progenitor, or even place meaningful limits on one. No point source is visible at the position of AT\,2021biy, and only diffuse emission is seen. In addition, we checked the images taken by \textit{WISE} from 2010 to 2020. The final released {\it WISE} images on 2020 May 30 record the status about half year before the major outburst of AT\,2021biy. However, there is no signal from a pre-nova outburst.

The {\it Hubble Space Telescope (HST)} data are more useful. We downloaded $2 \times 350$\,s $F814W$ and $2 \times 338$\,s $F606W$ images of NGC\,4631 taken on 2003 August 3 from the Mikulski Archive for Space Telescopes. We also briefly examined earlier data taken with WFPC2 in the $F336W$ and $F606W$ filters, but as these provided shallower images in the same filter as the ACS data ($F606W$) or the position of AT\,2021biy was on the very edge of a detector ($F336W$), they were not further pursued.

A deep ($15 \times 60$\,s) stack of $i$-band images taken with LT+IO:O on 2021 July 16 was used to identify AT\,2021biy in our pre-outburst data. The LT images were taken under good conditions, and the final stack has a FWHM of 0.9\arcsec. Ten point-like sources were identified in both the LT image and the pipeline-drizzled {\it HST}+ACS $F606W$ image. Using the matched coordinates of these sources in both frames, we derived a geometric transformation between the pre-outburst and post-outburst images with a root-mean-square scatter of only 0.09\arcsec.

\begin{figure}[htp]
\centering
\includegraphics[width=0.46\textwidth]{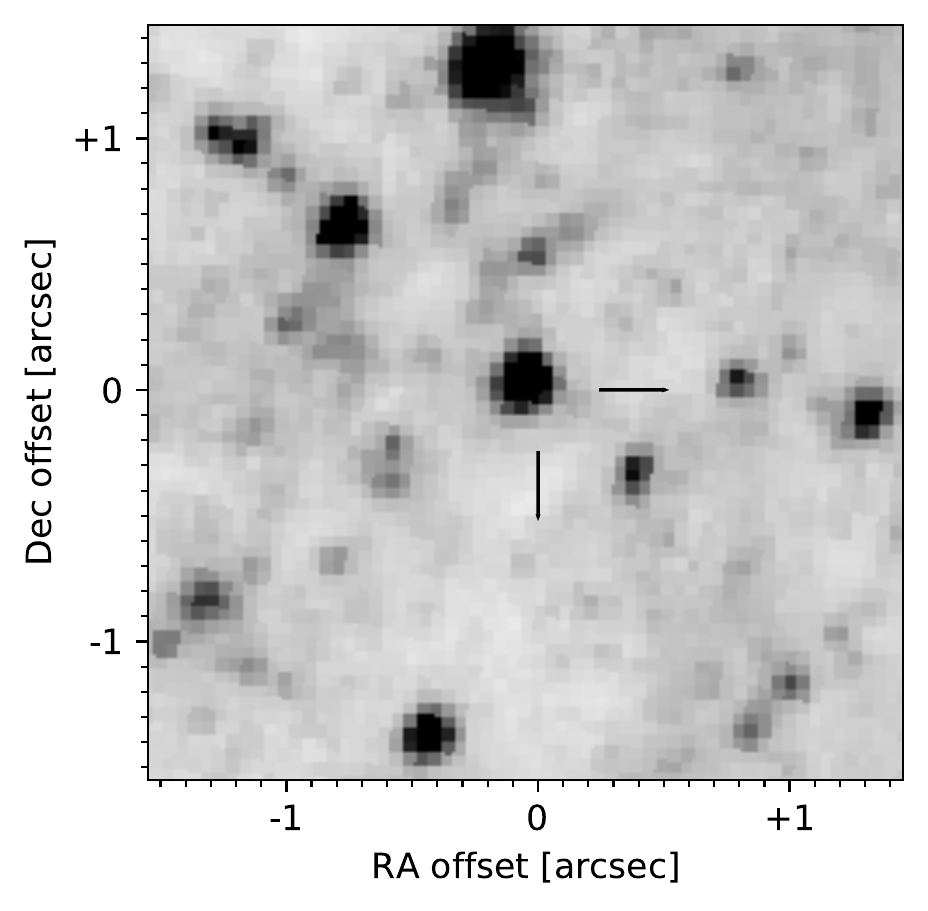}
\caption{{\it HST}+ACS $F606W$ image showing the location of AT\,2021biy, indicated with cross marks. A bright source is seen in the pre-explosion images coincident with the position of the transient.}
\label{21biyHST}
\end{figure}

We measured the position of AT\,2021biy in the LT image, and use our transformation to determine its corresponding position in the ACS image. A point source is visible in the ACS image with an offset of 0.08\arcsec\ from our transformed position (i.e. within the astrometric uncertainty). We show this source (henceforth referred to as the progenitor candidate) in Figure~\ref{21biyHST}.

We performed photometry on the {\it HST} images using {\sc Dolphot} \citep{Dolphin2016ascl.soft08013D}, and measured $F606W = 22.038 \pm 0.007$\,mag and $F814W = 21.408 \pm 0.008$\,mag for the progenitor in the Vega magnitude system. Given our adopted distance and reddening ($d = 7.46 \pm 0.50$\,Mpc; $E(B-V) = 0.271 \pm 0.096$\,mag), and converting to standard Johnson-Cousins filters, this corresponds to $M_V = -7.93 \pm 0.17$\,mag and $V-I = 0.52 \pm 0.01$\,mag. 

In order to determine the luminosity of this source, we use the YBC database \citep{Chen2019A&A...632A.105C}. We find that the colour and magnitude of the progenitor candidate match an ATLAS9 \citep{Castelli2003IAUS..210P.A20C} model at solar metallicity with log\,($L/{\rm L}_{\odot}$) = 5.0\,dex and $T_{\rm{eff}}  = 5900$\,K. Such a star would be a luminous yellow supergiant, and consistent with a zero-age main-sequence (ZAMS) mass of $\sim 20$\,\msun.

We also conducted a more systematic analysis of the progenitor candidate using the Hoki code \citep{Stevance2020JOSS....5.1987S} to explore a grid of BPASS models \citep{Eldridge2017PASA...34...58E, Stanway2018MNRAS.479...75S}. The BPASS model grid contains both single-star evolutionary tracks and binary models, making it especially suitable for analysing the progenitors of LRNe. We remark, however, that BPASS does not resolve the common envelope phase in the models on a dynamical timescale. We searched the grid of solar-metallicity BPASS models for any binary models that matched the measured progenitor magnitude in $F606W$ to within 0.17\,mag, and match the $F606W-F814W$ colour of the progenitor to within 0.05\,mag. We note that the tolerance in colour is larger than our photometric uncertainty in order to account for possible model systematics as well as the possibility that two consecutive model points may change by more than 0.01\,mag owing to sparse grid resolution.

\begin{figure*}[htbp]
\centering
\includegraphics[width=0.9\textwidth]{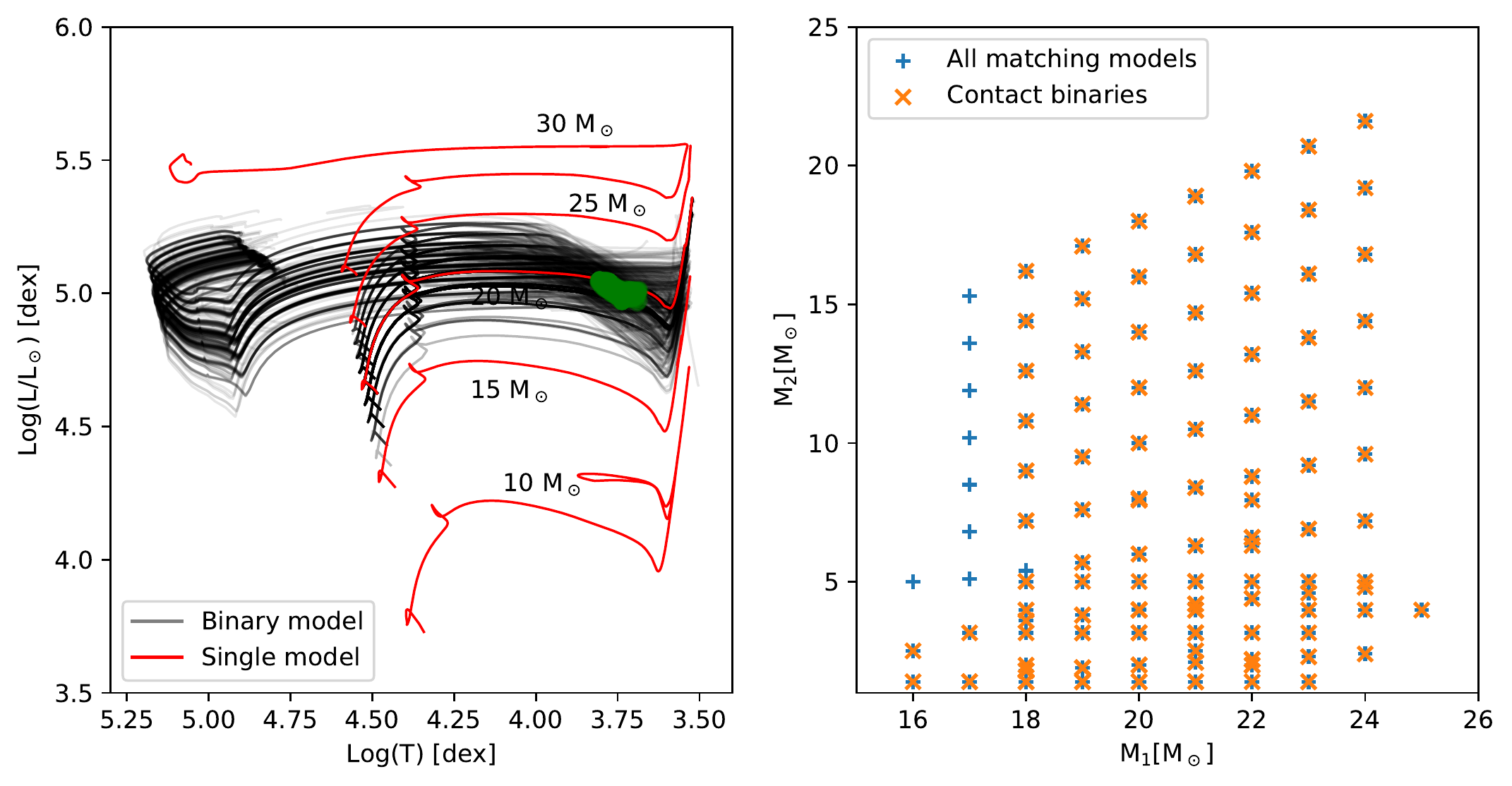}
\caption{Progenitor modelling of AT\,2021biy. {\it Left:} Hertzsprung-Russell diagram (HRD) for the progenitor candidate for AT\,2021biy. A set of single-star models is shown in red and labeled with their ZAMS mass, while binary models that match the progenitor magnitude and colour are shown as black lines. The point at which each model matches the observed progenitor photometry is marked in green. {\it Right:} The ZAMS mass of the primary and secondary for BPASS models that match the progenitor candidate.} 
\label{21biyHRD}
\end{figure*}

We plot the matching models in Figure~\ref{21biyHRD}. A large number of models  match our photometry, and these are also consistent with the luminosity inferred from the YBC database. We note that detailed MESA \citep[Modules for Experiments in Stellar Astrophysics;][]{Paxton2018ApJS..234...34P} modeling in \citet[][]{Blagorodnova2021A&A...653A.134B} for the common envelope phase in AT\,2018bwo shows a drop in luminosity. While a similar drop is also observed in many of our matching BPASS models, we caution that significant uncertainties exist in the treatment of the common envelope phase in BPASS (and probably in all stellar evolution codes). Encouragingly, we also see that a 20\,\msun\ single-star model from BPASS is consistent with the observed photometry. Turning to the binary models, we find that the majority of matching models have a primary with a ZAMS mass of 17--24\,\msun.
The mass of the secondary is less constrained. However, we note that weighting for the initial mass function (IMF), most models seem to have lower mass secondaries ($\sim60\%$ have M$_2$ below $\sim5$ M$_\odot$).

We also determined the binary separation for each of the matching models. Figure~\ref{21biyHRD} indicates models where the semimajor axis of the binary is less than the combined radius of the primary and secondary. We find that, again, the progenitor candidate is consistent with a close binary where the primary has  17--24\,\msun.

While the overall result from our {\it HST} data analysis is that the progenitor of AT\,2021biy is consistent with either a $\sim 20$\,\msun\ single star or a 17--24\,\msun\ ZAMS primary with a binary companion, we must stress several caveats. First, the progenitor candidate may be an erroneous identification. Attempting to perform differential astrometry between relatively wide-field, low-resolution, ground-based data and {\it HST} images with a small number of reference stars is subject to error. If the progenitor of AT\,2021biy is not the star we have identified, then it is necessarily fainter and likely has lower mass than our candidate.

We are also assuming that the progenitor magnitude has remained constant over the 18\,yr between the acquisition of the {\it HST}+ACS images and the discovery of AT\,2021biy. LRN progenitors have been seen to undergo a secular increase in brightness over timescales of years \citep[e.g. ][]{Tylenda2011A&A...528A.114T, Blagorodnova2017ApJ...834..107B}. If the progenitor was not in hydrostatic equilibrium in 2003, then the estimate of progenitor mass from comparison to the BPASS models is no longer valid (although it is probably still valid as an upper limit). Finally, we caution that we have neglected to account for circumstellar dust. With detections of the progenitor in only two bands, it is not possible to fit the reddening and temperature simultaneously.

\section{Discussion and concluding remarks}
\label{sec:discussion}

AT\,2021biy is one of the best-observed LRNe, with extensive optical and NIR datasets during the major outburst phase. Although AT\,2021biy shares a number of common properties with other LRNe, it shows an unprecedentedly long plateau, and a pronounced hump in the late-time (post-plateau) light curve.

The field of AT\,2021biy was imaged for $\sim 7.6$\,yr prior to the outburst by the ATLAS and Pan-STARRS surveys. However, a weak source is observed at the LRN location only from about $-$230 days to $-$200 days before the outburst onset. The subsequent observational gap lasts for about 165 days. For this reason, the timing of the onset of the common-envelope phase is not well constrained.

AT\,2021biy shows a very fast initial rise lasting $\sim 7$\,days, reaching the first blue peak at a luminosity of $1.6 \times 10^{41}$\,erg\,s$^{-1}$. This short-duration peak can be comfortably explained by the outflow of material ejected by the stellar core coalescence \citep[e.g.][]{Tylenda2006A&A...451..223T,Shara2010ApJ...725..831S,Nandez2014ApJ...786...39N,MacLeod2017ApJ...835..282M,Metzger2017MNRAS.471.3200M}. 
The first peak is followed by a rapid luminosity decline and then by a long plateau of $\sim 210$\,days. This is the longest plateau observed so far among LRNe. 

By analogy with SNe\,IIP, the plateau (or the equivalent second peak commonly observed in LRNe) is likely powered by hydrogen recombination, during which the largest fraction of the radiated energy in LRNe is emitted \citep[e.g. ][]{Ivanova2013Sci...339..433I, Ivanova2013A&ARv..21...59I, MacLeod2017ApJ...835..282M}. The recombination explanation is supported by several observed parameters: the minor evolution of temperature at $T \approx 4050$--4500\,K allows the H-rich gas to completely recombine (see Fig.~\ref{pic:SEDs}); the \Ha\ emission line becomes quite weak, and at the same time a forest of metal lines in absorption appears (see Figs.~\ref{spectseq} and \ref{Hprofiles}), as in some low-luminosity SNe\,IIP \citep[e.g. ][]{Pastorello2004MNRAS.347...74P, Pastorello2006MNRAS.370.1752P, Pastorello2009MNRAS.394.2266P, Spiro2014MNRAS.439.2873S, Hosseinzadeh2018ApJ...861...63H, MullerBravo2020MNRAS.497..361M, Reguitti2021MNRAS.501.1059R, Valerin2022MNRAS.513.4983V, Zhang2022MNRAS.513.4556Z}. 

As discussed by \citet[][]{Matsumoto2022arXiv220210478M}, the plateau duration is mostly determined by the timescale of the H recombination, which occurs at $T \approx T_{\mathrm{i}}$. With this $T_{\mathrm{i}}$, the characteristic density ($\rho_{\mathrm{i}}$) is approximately a constant of $10^{-11}$\,g\,cm$^{-3}$ in the Saha equation. Therefore, the duration of an LRN plateau is set by $\rho_{\mathrm{i}}$ and other parameters, following the relation

\begin{equation} \label{ejecta}
    t_{\mathrm{pl}} \approx \left(\frac{3M_{\mathrm{ej}}}{4\pi \rho_{\mathrm{i}} \overline{v}^3_\mathrm{E}}\right)^{1/3} \approx 140\,d\,\rho_{\mathrm{i,-11}}^{-1/3} 
    \left(\frac{M_{\mathrm{ej}}}{\msun}\right)^{1/3} 
    \left( \frac{\overline{v}_\mathrm{E}}{300\,\kms} \right)^{-1}\, ,
\end{equation}
\\
where $\rho_{\mathrm{i,-11}} = \rho_{\mathrm{i}}/10^{-11}$\,g\,cm$^{-3}$, $M_{\mathrm{ej}}$ is the total ejecta mass, and $\overline{v}_\mathrm{E}$ is the mean ejecta velocity. Here, we estimate $M_{\mathrm{ej}}$ following Eq.~\ref{ejecta}. In the case of AT\,2021biy, the plateau duration is obtained from the light curve with $t_{\mathrm{pl}} = 210$\,days, and the ejecta velocity ($\overline{v}_\mathrm{E} = 430$\,\kms) is estimated through the FWHM velocity of H$\alpha$ in the $+$177.2\,day Shane/Kast spectrum taken during the plateau phase (see Sect. \ref{specevol}). We find $M_{\mathrm{ej}} = 9.9$\,\msun. Note that this result should be considered only as an indicative value, because the above relation is highly sensitive to the ejecta velocity which was estimated through our moderate-resolution spectrum. This should be improved by future efforts in both theoretical modeling and high-resolution spectral observations.  Furthermore, we follow \citet[][]{Matsumoto2022arXiv220210478M} to infer the luminosity at plateau phase (see their Eq. 18). 
Adopting the above parameters ($\overline{v}_\mathrm{E} = 430$\,\kms~and $M_{\mathrm{ej}} = 9.9$\,\msun), the plateau luminosity results to be $L_{\mathrm{pl}} \approx~3 \times 10^{39}$\,erg\,s$^{-1}$, which is about one order of magnitude smaller than the observed one ($\sim 5 \times 10^{40}$\,erg\,s$^{-1}$, see the top-right panel of Fig.~\ref{pic:SEDs}). This discrepancy  can be attributable to the crude assumption that the LRN plateau is solely powered by H recombination \citep[][]{Matsumoto2022arXiv220210478M}. This tension can be relaxed by invoking an extra heating source, such as embedded shock-interaction between the ejecta and the circumbinary material \citep[e.g. ][]{Metzger2017MNRAS.471.3200M}. 

Similar to other LRNe, the plateau of AT\,2021biy is followed by a rapid luminosity decline. However, the decline abruptly stopped 11\,months after maximum brightness, and the light curve showed a short-lived hump in all bands, likely due to gas shell collisions. Similar short-lived fluctuations in the light curve are frequently observed in some SNe whose ejecta interact with circumstellar material, and are normally interpreted as ejecta collisions with relatively thin circumstellar shells \citep[e.g.][]{Pastorello2008MNRAS.389..131P,Pastorello2015MNRAS.449.1921P,Smith2012MNRAS.426.1905S,Martin2015AJ....149....9M,Reguitti2019MNRAS.482.2750R}.

From the spectroscopic point of view, AT\,2021biy shows the canonical evolution of LRNe. Early-time spectra are characterised by a blue continuum with prominent H lines. During the plateau phase, the spectra become redder and are dominated by a forest of narrow metal lines in absorption. 
Finally, all spectra at late times are extremely red and with prominent molecular features, resembling those of M-type stars.

LRNe are expected to produce dusty environments \citep[e.g. ][]{Nicholls2013MNRAS.431L..33N, Zhu2013ApJ...777...23Z, Goranskij2020AstBu..75..325G, Mobeen2021A&A...655A.100M, Woodward2021AJ....162..183W}. 
Spectropolarimetric analysis reveals that the size distribution of the interstellar dust grains in the vicinity of AT\,2021biy resembles that around V838\,Mon in the Milky Way.

Many authors suggest that physical parameters of LRNe are correlated \citep[e.g. ][]{Pejcha2016MNRAS.455.4351P,Mauerhan2018MNRAS.473.3765M,Pastorello2019A&A...630A..75P,Pastorello2021A&A...647A..93P,Blagorodnova2021A&A...653A.134B}. In particular, the light curve luminosity seems to be correlated with some spectral parameters, such as the \Ha~luminosity ($L_{\rm{\Ha}}$) and the \Ha~FWHM velocity ($v_{\rm{FWHM}}$). They inferred a general trend according to which brighter events have higher $L_{\Ha}$  and $v_{\rm{FWHM}}$(\Ha) \citep[see the analysis in ][ and their Fig. 17]{Pastorello2022arXiv220802782P}.

\begin{figure}[htbp]
\centering
\includegraphics[width=0.48\textwidth]{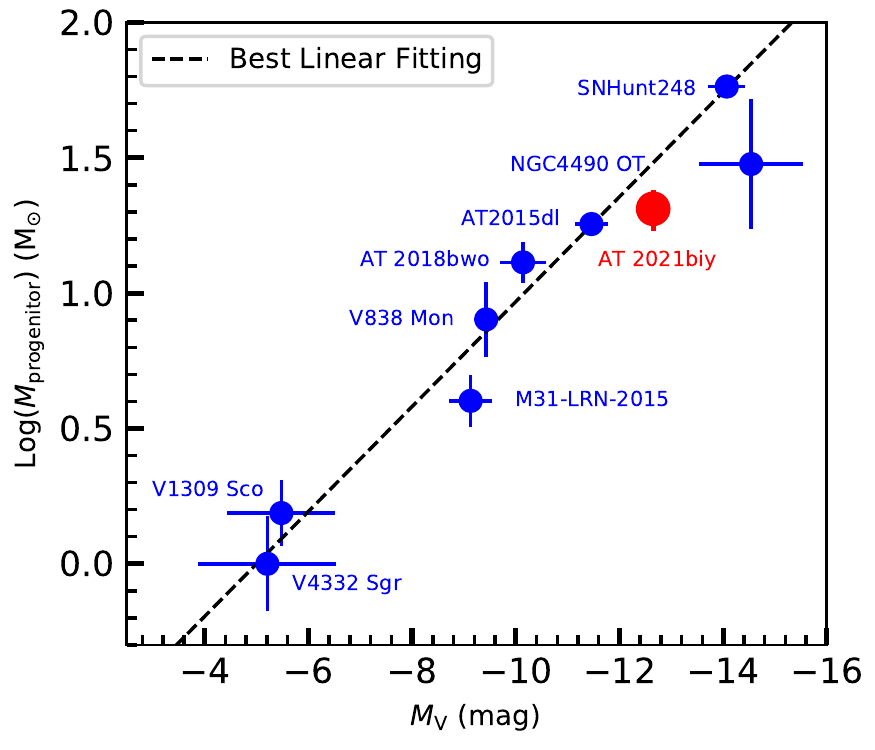}
\caption{Diagram of progenitor mass (log\,$(M_{\mathrm{progenitor}})$) versus $V$-band second peak (or plateau) magnitude ($M_V$). Blue dots represent data from the literature, while the red point marks the location of AT\,2021biy. The linear best fit to the data is shown as a dashed line.
Data for LRNe are from from the literature \citep[][]{Martini1999AJ....118.1034M,Munari2002A&A...389L..51M,Mason2010A&A...516A.108M,Tylenda2011A&A...528A.114T,Kankare2015A&A...581L...4K,Mauerhan2015MNRAS.447.1922M,Williams2015ApJ...805L..18W, Smith2016MNRAS.458..950S, Blagorodnova2017ApJ...834..107B,Kaminski2018A&A...617A.129K,Pastorello2019A&A...630A..75P,Ortiz2020A&A...638A..17O,Blagorodnova2021A&A...653A.134B}.
}
\label{MassMag}
\end{figure}

Furthermore, the luminosity of the light curve is tightly correlated with that of the progenitor. \cite{Kochanek2014MNRAS.443.1319K} and \cite{Blagorodnova2021A&A...653A.134B} determined that more-massive progenitor systems are likely to produce more-luminous LRN outbursts. In particular, \cite{Blagorodnova2021A&A...653A.134B} presented a tentative progenitor mass vs. peak luminosity diagram showing a robust correlation between these two parameters. In Figure~\ref{MassMag}, we update that diagram using the same sample as \cite{Blagorodnova2021A&A...653A.134B}, and including AT\,2021biy.

We searched the literature for the masses of the quiescent progenitors of all LRNe for which that parameter was estimated, while for consistency we measured the $V$-band absolute magnitude ($M_V$) at the red peak (or in the plateau for those objects that do not show a second peak) for all objects\footnote{We note that the luminosity of the first peak depends on many factors such as the expansion and the radius of the primary star, while the second peak/plateau depends on the mass of the recombining hydrogen (which is likely correlated with the total systemic mass). Therefore, we use the magnitude at the second peak/plateau as the reference.}. The absolute magnitude and progenitor mass values we used are reported in Table \ref{parameters_relation_lrne} \cite[see][for the relative references]{Pastorello2022arXiv220802782P}. Even though for only 9 objects\footnote{Within this LRN sample, SNHunt248 should be regarded as an LRN candidate; its nature is still debated \citep{Kankare2015A&A...581L...4K}.} the progenitor mass was well estimated from pre-explosion images or modeling of the light curves, we see a clear linear correlation between these two parameters.
We applied the Pearson test to evaluate the strength of the correlation between log\,$(M_{\mathrm{progenitor}})$ and $M_V$, obtaining a coefficient of $r=-0.97$ that indicates a very strong negative correlation.
Therefore, we also performed a linear regression on the data weighted by the uncertainties in the absolute magnitudes, deriving the relation
\begin{equation} \label{equation2}
    log\,(M/M_{\odot}) = (-0.162 \pm 0.020)\,M_V \textrm{(mag)} - (0.701 \pm 0.048)\, .
\end{equation}
The above expression can be also written as
\begin{equation}
    M_V=(-5.56\pm0.69)\,{\rm log}(M/M_{\odot}) - (4.91 \pm 0.26)~\textrm{mag} \, .
\end{equation}
In terms of luminosity, it is equivalent to $L\propto M^{2.22\pm0.28}$.
For reference, the original relations found by \cite{Kochanek2014MNRAS.443.1319K} were $M_V=(-7.2\pm1.5) \times {\rm log}(M/{\rm M}_{\odot}) - (3.6\pm0.8)$\,mag and $L\propto M^{2.9\pm0.6}$.
The expanding sample of LRNe with progenitor information in the future will help refine the relation.

AT\,2021biy is one of the LRNe with the best-constrained progenitor. The detection of a source at the LRN location in archival {\it HST} images allows us to constrain its progenitor to be either a luminous yellow supergiant or, most likely, a binary system with a primary of 17--24\,\msun, placing AT\,2021biy in the small group of LRNe with very massive progenitors, although still a factor of 2--3 less than the most massive progenitor of an LRN observed so far for SNHunt 248 \citep{Mauerhan2018MNRAS.473.3765M}.

Future facilities, such as the China Space Station Telescope \citep[CSST; ][]{Cao2022MNRAS.511.1830C,Cao2022RAA....22b5019C}, the Vera C. Rubin Observatory \citep[e.g. ][]{Li2022ApJS..258....2L,Bianco2022ApJS..258....1B}, and the Nancy Grace Roman Space Telescope \citep[e.g. ][]{Rubin2021PASP..133f4001R, Gezari2022arXiv220212311G} will be essential for substantially increasing the population of LRNe. Joint observational and theoretical efforts will allow us to place tight constraints on the nature of LRNe.

\section*{Acknowledgments}
The full acknowledgments are available in Appendix \ref{App:Acknowledgements}.

\bibliographystyle{aa} 
\bibliography{gap}

\begin{appendix}
\onecolumn

\section{Supplementary materials}
\label{sec:appendix}

\subsection{Photometric data for AT\,2021biy}
\label{sec:photdata}
Optical and NIR photometric measurements of AT\,2021biy are only available in electronic form at the CDS via anonymous ftp to \url{cdsarc.u-strasbg.fr (130.79.128.5)} or via \url{http://cdsweb.u-strasbg.fr/cgi-bin/qcat?J/A+A/}, while other photometric materials are reported here. Our observations will be made public via the Weizmann Interactive Supernova Data Repository \citep[WISeREP;][]{Yaron2012PASP..124..668Y}.\\ 


\begin{table}[!h]
\centering
\caption{Decline rates of the light curves of AT\,2021biy (mag (100\,day)$^{-1}$) with their uncertainties.}
\label{lc_para}
	\scalebox{1.}{
		\begin{tabular}{@{}ccccccc@{}}
		\hline
		Filter & Phase 1 ($\gamma_1$) & Phase 2 ($\gamma_2$) & Phase 3 ($\gamma_3$) & Phase 4 ($\gamma_4$) & Phase 5 ($\gamma_5$)& Phase 6 ($\gamma_6$)\\
		\hline
		  & $\sim 0$--50\,d & $\sim 50$--100\,d & $\sim 100$--310\,d & $\sim 310$--340\,d & $\sim 340$--370\,d & $\sim 370$--430\,d \\  
		\hline
		$u$  & 10.87 (2.54)& --             &  --          &  --          & --             & --\\
		$B$  & 6.90 (0.36) & $-$1.70 (0.13) & 0.17 (0.02)  &  8.82 (0.84) & --             & --\\
		$g$  & 6.37 (0.29) & $-$1.97 (0.08) & 0.13 (0.01)  &  8.13 (1.02) & $-$2.27 (0.06) & 6.15 (0.12)\\
		$c$  & 4.69 (0.55) & $-$1.75 (0.30) & 0.28 (0.07)  &  --          & --             & 5.72 (1.55)\\
		$V$  & 4.55 (0.30) & $-$1.50 (0.09) & 0.21 (0.03)  &  7.25 (1.15) & $-$2.23 (0.51) & --\\
		$r$  & 4.58 (0.29) & $-$1.67 (0.06) & 0.20 (0.02)  & 7.94 (0.70)  & $-$2.25 (0.35) & 5.58 (0.48)\\
		$o$  & 3.52 (0.83) & $-$1.64 (0.07) & 0.18 (0.02)  & 7.51 (0.23)  & $-$2.35 (0.23) & 3.39 (0.15)\\
		$i$  & 4.79 (0.41) & $-$1.69 (0.09) & 0.10 (0.01)  & 6.19 (0.61)  & $-$2.26 (0.27) & 5.29 (0.30)\\
		$z$  & 2.69 (0.37) & $-$1.22 (0.07) & 0.13 (0.05)  & 5.76 (0.47)  & $-$1.58 (0.15) & 4.18 (0.33)\\
		$y/Y$& --          &  --            & 0.06 (0.03)  & 3.47 (0.41)  & $-$1.99 (0.16) & 2.24 (0.60)\\
		$J$  & --          & $-$1.09 (0.18) & 0.01 (0.04)  & 2.78 (0.39)  & $-$0.59 (0.29) & 1.47 (0.17)\\
		$H$  & --          & $-$1.08 (0.11) & 0.01 (0.06)  & 2.08 (0.32)  & $-$0.22 (0.14) & 1.04 (0.14)\\
		$K$  & --          & $-$1.23 (0.23) & 0.02 (0.03)  & 1.43 (0.24)  & $-$0.14 (0.09) & 0.63 (0.10)\\
		\hline
		\end{tabular}
	}
	\medskip
\end{table}

\begin{table*}
\centering
	\caption{Parameters of black-body fit to the $uBgcVroizyYJHK$ bands of AT\,2021biy. Uncertainties are given in parentheses.}
	\label{AT2021biySED}
	\begin{tabular}{@{}cccccc@{}}
		\hline
		Date & MJD & Phase$^a$  &   Luminosity    &  Temperature    &    Radius   \\ 
		     &     & (days)  &  ($10^{40}$\,erg\,s$^{-1}$) &  (K)   &   ($10^{14}$\,cm) \\
		\hline
		20210131 & 59245.10 & $-5.9$  & 4.97 (2.17) &  9450 (590)  &   0.94 (0.20) \\
		20210202 & 59247.17 & $-3.8$  & 9.78 (4.03) &  9615 (565)  &   1.27 (0.26) \\
		20210203 & 59248.20 & $-2.8$  & 11.50 (4.53)&  10080 (565) &   1.25 (0.25) \\
		20210205 & 59250.45 & $-0.6$  & 15.91 (13.04)& 11430 (1410) &  1.14 (0.47) \\
		20210210 & 59255.09 & $+4.1$  & 12.54 (8.91)&  10305 (1055) &  1.25 (0.44) \\
		20210216 & 59261.12 & $+10.1$ & 7.66 (1.80) &  9190 (260)  &   1.23 (0.14) \\
		20210219 & 59264.80 & $+13.8$ & 5.78 (1.81) &  8685 (395)  &   1.19 (0.19) \\
		20210227 & 59272.11 & $+21.1$ & 3.21 (0.92) &  6630 (225)  &   1.53 (0.28) \\
		20210303 & 59276.08 & $+25.1$ & 2.49 (0.88) &  5980 (255)  &   1.65 (0.29) \\
		20210313 & 59285.99 & $+35.0$ & 2.16 (0.59) &  5255 (145)  &   1.99 (0.27) \\
		20210319 & 59292.05 & $+41.0$ & 2.02 (0.62) &  4950 (155)  &   2.17 (0.33) \\
		20210327 & 59300.43 & $+49.4$ & 2.06 (0.76) &  4495 (170)  &   2.66 (0.49) \\
		20210403 & 59307.98 & $+57.0$ & 1.86 (0.48) &  4780 (140)  &   2.24 (0.29) \\
		20210414 & 59318.99 & $+68.0$ & 2.32 (1.16) &  4585 (250)  &   2.71 (0.68) \\
		20210429 & 59333.10 & $+82.1$ & 2.76 (0.66) &  4655 (115)  &   2.87 (0.34) \\
		20210508 & 59342.01 & $+91.0$ & 3.48 (2.00) &  4630 (280)  &   3.26 (0.92) \\
		20210519 & 59353.91 & $+102.9$& 4.41 (1.43) &  4555 (155)  &   3.79 (0.61) \\
		20210530 & 59364.98 & $+114.0$& 4.94 (2.06) &  4325 (190)  &   4.45 (0.93) \\
		20210616 & 59381.93 & $+130.9$& 5.07 (1.78) &  4145 (145)  &   4.91 (0.86) \\
		20210626 & 59391.22 & $+140.2$& 4.67 (1.19) &  4340 (155)  &   4.05 (0.52) \\
		20210709 & 59404.89 & $+153.9$& 4.96 (1.19) &  4355 (100)  &   4.40 (0.53) \\
		20210725 & 59420.85 & $+169.8$& 4.43 (0.91) &  4340 (80)   &   4.18 (0.43) \\
		20210808 & 59434.88 & $+183.9$& 4.18 (0.89) &  4400 (95)   &   3.96 (0.42) \\
		20210820 & 59446.89 & $+195.9$& 3.97 (0.85) &  4360 (95)   &   3.92 (0.42) \\
		20210911 & 59468.78 & $+217.8$& 4.17 (0.82) &  4230 (75)   &   4.27 (0.42) \\
		20211105 & 59523.16 & $+272.2$& 4.50 (1.10) &  4090 (90)   &   4.74 (0.58) \\
		20211202 & 59550.48 & $+299.5$& 4.31 (0.88) &  4185 (75)   &   4.44 (0.45) \\
		20211218 & 59566.18 & $+315.2$& 3.87 (0.84) &  4055 (80)   &   4.48 (0.49) \\
		20211225 & 59573.17 & $+322.2$& 3.04 (0.86) &  3590 (100)  &   5.07 (0.71) \\ 
		20220102 & 59581.19 & $+330.2$& 1.93 (0.71) &  3090 (105)  &   5.44 (1.00) \\ 
		20220113 & 59592.11 & $+341.1$& 1.88 (1.23) &  2825 (165)  &   6.44 (2.11) \\  
		20220123 & 59602.26 & $+351.3$& 1.89 (0.56) &  3030 (80)   &   5.62 (0.83) \\ 
		20220204 & 59614.18 & $+363.2$& 2.26 (0.53) &  3140 (65)   &   5.71 (0.67) \\ 
		20220212 & 59622.37 & $+371.4$& 2.18 (0.45) &  3170 (50)   &   5.50 (0.56) \\ 
		20220214 & 59624.02 & $+373.0$& 2.16 (0.48) &  3105 (55)   &   5.71 (0.64) \\ 
		20220228 & 59638.23 & $+387.2$& 1.91 (0.45) &  2815 (55)   &   6.53 (0.78) \\ 
		20220409 & 59678.90 & $+427.9$& 1.11 (0.58) &  2055 (95)   &   9.35 (2.43) \\ 
        20220414 & 59683.98 & $+433.0$& 0.87 (0.47) &  2380 (140)  &   6.16 (1.66) \\ 
        20220424 & 59693.93 & $+442.9$& 0.98 (0.38) &  2330 (90)   &   6.83 (1.31) \\ 
        20220518 & 59717.94 & $+466.9$& 0.85 (0.41) &  2300 (110)  &   6.54 (1.58) \\  
		\hline
	\end{tabular}
	\medskip
	
	$^a$Phases are relative to $r$-band maximum brightness (MJD = 59251.0 $\pm$ 1.0).\\
\end{table*}

\onecolumn
\subsection{Log of the spectroscopic observations of AT\,2021biy.}
\label{app:speclog}
\begin{table*}[h]
\caption{General information on the spectroscopic observations of AT\,2021biy.}
\label{2021biyspecinfo}
\begin{tabular}{@{}cccccccc@{}}
\hline
Date & MJD & Phase$^a$ & Telescope+Instrument & Grism+Slit & Spectral range & Resolution & Exp. time \\ 
  &   & (days) &   &        & (\AA)    & (\AA)           & (s)           \\ 
\hline
20210130 & 59244.45 & $-$6.6   &  OGG 2m+FLOYDS   & red/blu+2.0"         & 3500-10000  &  13   & 2700       \\
20210131 & 59245.12 & $-$5.9   &  NOT+ALFOSC      & gm4+1.0"             & 3400-9680   &  14   & 2700       \\
20210201 & 59246.48 & $-$4.5   &  OGG 2m+FLOYDS   & red/blu+2.0"         & 3500-10000  &  13   & 3600       \\
20210201 & 59246.77 & $-$4.2   &  LJT+yf01        & grism3+2.0"          & 3500-8740   &  21   & 2000       \\
20210203 & 59248.73 & $-$2.3   &  LJT+yf01        & grism3+2.0"          & 3500-8740   &  21   & 2000       \\
20210207 & 59252.46 & $+$1.5   &  Shane+Kast      & 300/7500+2.0"        & 3620-10700  &  9.3  & 2160/2100  \\
20210210 & 59255.68 & $+$4.7   &  LJT+yf01        & grism3+2.0"          & 3500-8740   &  21   & 2200       \\
20210214 & 59259.86 & $+$8.9   &  DOT+ADFOSC      & GR676R+1.5"          & 3800-8900   &  12   & 900        \\
20210216 & 59261.13 & $+$10.1  &  Ekar1.82m+AFOSC & VPH6+1.69"           & 5000-9290   &  16   & 2400       \\
20210218 & 59263.41 & $+$12.4  &  Shane+Kast      & 300/7500+2.0"        & 3630-10740  &  9.3  & 2160/2100  \\
20210307 & 59280.52 & $+$29.5  &  Shane+Kast      & 300/7500+2.0"        & 3620-10730  &  9.3  & 3060/3000  \\
20210323 & 59296.02 & $+$45.0  &  NOT+ALFOSC      & gm4+1.0"             & 3400-9680   &  14   & 1800       \\
20210403 & 59307.95 & $+$57.0  &  NOT+ALFOSC      & gm4+1.0"             & 3400-9680   &  14   & 3600       \\
20210407 & 59311.53 & $+$60.5  &  Keck-I+LRIS  & 600/4000\&400/8500+1.0" & 3130-10280  & 4.7/9.0 & 900/900  \\
20210418 & 59322.26 & $+$71.3  &  Shane+Kast      & 300/7500+2.0"        & 3630-10740  &  9.3  & 3060/3000  \\
20210518 & 59352.53 & $+$101.5 &  Shane+Kast      & 300/7500+2.0"        & 3630-10750  &  9.3  & 3360/3300  \\
20210615 & 59380.90 & $+$129.9 &  BTA+SCORPIO-1   & VPHG1200R+1.2"       & 5660-7375   &  5.6  & 2700   \\
20210615 & 59380.99 & $+$130.0 &  NOT+ALFOSC      & gm4+1.0"             & 3400-9650   &  14   & 2700   \\
20210703 & 59398.26 & $+$147.3 &  Shane+Kast      & 300/7500+2.0"        & 3630-10750  &  9.3  & 3660/3600  \\
20210705 & 59400.96 & $+$150.0 &  TNG+LRS         & LRB/LRR+1.0"         & 3500-10370  &  11/10 & 1200/1200  \\
20210716 & 59411.19 & $+$160.2 &  Shane+Kast      & 300/7500+2.0"        & 3630-10750  &  9.3   & 3660/3600  \\
20210717 & 59412.91 & $+$161.9 &  GTC+OSIRIS      & R1000B/R1000R+1.0"   & 3640-10140  &  7/8   & 1500/1500  \\
20210802 & 59428.18 & $+$177.2 &  Shane+Kast      & 300/7500+2.0"        & 3630-10750  &  9.3   & 3660/3600  \\
20210819 & 59445.88 & $+$194.9 &  NOT+ALFOSC      & gm4+1.0"             & 3400-9650   &  14    & 2400   \\
20211103 & 59521.52 & $+$270.5 &  Shane+Kast      & 300/7500+2.0"        & 3630-10750  &  9.3   & 3660/3600  \\
20211104 & 59522.07 & $+$271.1 &  BTA+SCORPIO-2   & VPHG1200@540+1.0"    & 3800-7330   &  4     & 3600   \\
20211205 & 59553.53 & $+$302.5 &  Shane+Kast      & 300/7500+2.0"        & 3630-10750  &  9.3   & 4890/4800 \\
20211214 & 59562.10 & $+$311.1 &  BTA+SCORPIO-2   & VPHG1200@540+1.0"    & 3800-7330   &  4     & 4200  \\
20211228 & 59577.14 & $+$326.1 &  NOT+ALFOSC      & gm4+1.0"             & 3520-9360   &  14    & 3600  \\
20220110 & 59589.69 & $+$338.7 &  Keck-I+MOSFIRE    & JHK+0.7"     & 9000-25000  &  $R$=3270      & 120/120/120 \\
20220212 & 59622.09 & $+$371.1 &  GTC+EMIR        & YJ/HK+1.3"           & 9000-25000  & $R$=987 & 1440/2880  \\
20220213 & 59623.52 & $+$372.5 &  LBT+MODS        & G400L/G670L+1.0"     & 3500-10000  &  3.3   & 5$\times$600 \\
20220304 & 59642.60 & $+$391.6 &  Keck-I+LRIS  & 600/4000\&400/8500+1.0" & 3130-10280  & 4.7/9.0 & 900/900 \\
\hline
\end{tabular}

\medskip
$^a$Phases are relative to $r$-band maximum light (MJD = 59251.0 $\pm$ 1.0).
\end{table*}

\subsection{Parameters of the sample of LRNe}

\begin{table}[htbp]
\centering
\caption{$M_V$ at second peak (or plateau) and progenitor mass used to derive Eq.~\ref{equation2}.$^a$}
\label{parameters_relation_lrne}
	\begin{tabular}{ccc}
	\hline
	Name & Progenitor mass & $M_V$ \\
    	 & (\msun) & (mag) \\
	\hline
	V4332\,Sgr & 1.00 $\pm$ 0.5 & $-5.21 \pm 1.33$ \\
	V1309\,Sco & 1.54 $\pm$ 0.5 & $-5.48 \pm 1.04$ \\
	M31-LRN2015 & 4 $\pm$ 1 & $-9.13 \pm 0.42$ \\
	V838\,Mon     & 8 $\pm$ 3 & $-9.43 \pm 0.22$ \\
	AT\,2018bwo   & 13 $\pm$ 2.5 & $-10.14 \pm 0.45$ \\
	AT\,2015dl    & 18 $\pm$ 1 & $-11.46 \pm 0.31$ \\
	AT\,2021biy & 20.5 $\pm$ 3.5 & $-12.65 \pm 0.16$ \\
	NGC\,4490OT & 30 $\pm$ 22 & $-14.54 \pm 1.00$ \\
	SNhunt248 & 58 $\pm$ 2 & $-14.07 \pm 0.36$ \\
	\hline
	\end{tabular}
	\medskip
	
$^a${See \citet[][]{Pastorello2022arXiv220802782P} for references for the values of each object.}
\end{table}

\newpage
\section{Acknowledgements}\label{App:Acknowledgements}
We gratefully thank the anonymous referee for his/her insightful comments that improved the paper.
YZC thanks Z. Chen (THU-DoA), E. Mason, T. Nagao, A. Schultz, and C. Pellegrino for some observations and discussions. WC thanks Prof. Joseph Hennawi for helpful advice on the use of PypeIt telluric-correction scripts.
YZC is funded by China Postdoctoral Science Foundation (grant 2021M691821).
This work is supported by the National Natural Science Foundation of China (NSFC grants 12033003 and 11633002), the Scholar Program of Beijing Academy of Science and Technology (DZ:BS202002), and the Tencent Xplorer Prize. 
MF is supported by a Royal Society -- Science Foundation Ireland University Research Fellowship. 
SB is partially supported by PRIN-INAF 2017 of Toward the SKA and CTA era: discovery, localisation, and physics of transient sources. EC, NER, and LT acknowledge support from MIUR, PRIN 2017 (grant 20179ZF5KS) `The new frontier of the Multi-Messenger Astrophysics: follow-up of electromagnetic transient counterparts of gravitational wave sources.' 
NER also acknowledges partial support from the Spanish MICINN grant PID2019-108709GB-I00 and FEDER funds, and from the program Unidad de Excelencia Mar\'ia de Maeztu CEX2020-001058-M.
LG acknowledges financial support from the Spanish Ministerio de Ciencia e Innovaci\'on (MCIN), the Agencia Estatal de Investigaci\'on (AEI) 10.13039/501100011033, and the European Social Fund (ESF) `Investing in your future' under the 2019 Ram\'on y Cajal program RYC2019-027683-I and the PID2020-115253GA-I00 HOSTFLOWS project, from Centro Superior de Investigaciones Cient\'ificas (CSIC) under the PIE project 20215AT016, and the program Unidad de Excelencia Mar\'ia de Maeztu CEX2020-001058-M.
TR acknowledges the financial support of the Jenny and Antti Wihuri and the Vilho, Yrj{\"o} and Kalle V{\"a}is{\"a}l{\"a} foundations. 
AR acknowledges support from ANID BECAS/DOCTORADO NACIONAL 21202412.
MDS is supported by grants from the VILLUM FONDEN (grant 28021) and the Independent Research Fund Denmark (IRFD; 8021-00170B).
PLK is supported by U.S. National Science Foundation (NSF) grant AST-1908823.
RD acknowledges funds by ANID grant FONDECYT Postdoctorado \#3220449.
JJZ is supported by the NSFC (grants 11403096 and 11773067), the Key Research Program of the CAS (grant KJZD-EW-M06), the Youth Innovation Promotion Association of the CAS (grant 2018081), and the CAS `Light of West China' program.
AVF's group at UC Berkeley has received support from the Miller Institute for Basic Research in Science (where AVF was a Miller Senior Fellow), the Christopher R. Redlich Fund, numerous individual donors, and NASA/{\it HST} grant AR-14295 from the Space Telescope Science Institute (STScI), which is operated by the Association of Universities for Research in Astronomy (AURA), Inc., under NASA contract NAS5-26555.
This work is partially supported by the China Manned Spaced Project (CMS-CSST-2021-A12).
HG acknowledges support from NSFC (grant 12173081), the key research program of frontier sciences, CAS, \#ZDBS-LY-7005, and Yunnan Fundamental Research Projects (grant 202101AV070001).
LX is thankful for support from the National Natural Science Foundation of China (grant 12103050 ), Advanced Talents Incubation Program of the Hebei University, and Midwest Universities Comprehensive Strength Promotion project.
SM acknowledges support from the Magnus Ehrnrooth Foundation and the Vilho, Yrj{\"o} and Kalle V{\"a}is{\"a}l{\"a} Foundation.
KM acknowledges BRICS grant DST/IMRCD/BRICS/Pilotcall/ProFCheap/2017(G) for this work. 
PJV is supported by the U.S. National Science Foundation Graduate Research Fellowship Program under grant DGE-1343012.
The research of Y.\ Yang has been supported through the Bengier-Winslow-Robertson Fellowship.
We acknowledge the support of the staffs of the various observatories at which data were obtained.
Funding for the LJT has been provided by Chinese Academy of Sciences and the People's Government of Yunnan Province. The LJT is jointly operated and administrated by Yunnan Observatories and Center for Astronomical Mega-Science, CAS.
Based on observations made with the Nordic Optical Telescope, owned in collaboration by the University of Turku and Aarhus University, and operated jointly by Aarhus University, the University of Turku, and the University of Oslo, representing Denmark, Finland, and Norway, the University of Iceland, and Stockholm University at the Observatorio del Roque de los Muchachos, La Palma, Spain, of the Instituto de Astrofisica de Canarias.
Observations from the NOT were obtained through the NUTS2 collaboration which is supported in part by the Instrument Centre for Danish Astrophysics (IDA). The data presented here were obtained in part with ALFOSC, which is provided by the Instituto de Astrofisica de Andalucia (IAA) under a joint agreement with the University of Copenhagen and NOTSA.
The Liverpool Telescope is operated on the island of La Palma by Liverpool John Moores University in the Spanish Observatorio del Roque de los Muchachos of the Instituto de Astrofisica de Canarias with financial support from the UK Science and Technology Facilities Council.
The Italian Telescopio Nazionale Galileo (TNG) operated on the island of La Palma by the Fundaci\'on Galileo Galilei of the INAF (Istituto Nazionale di Astrofisica) at the Spanish Observatorio del Roque de los Muchachos of the Instituto de Astrofísica de Canarias.
Based on observations obtained with the Gran Telescopio Canarias (GTC), installed in the Spanish Observatorio del Roque de los Muchachos of the Instituto de Astrofisica de Canarias, on the island of La Palma.
Based on observations collected at Copernico and Schmidt telescopes (Asiago, Italy) of the INAF -- Osservatorio Astronomico di Padova.
The W. M. Keck Observatory is operated as a scientific partnership among the California Institute of Technology, the University of California and NASA; the observatory was made possible by the generous financial support of the W. M. Keck Foundation. A major upgrade of the Kast spectrograph on the Shane 3\,m telescope at Lick Observatory was made possible through generous gifts from William and Marina Kast as well as the Heising-Simons Foundation. Research at Lick Observatory is partially supported by a generous gift from Google.
This work makes use of data from the Las Cumbres Observatory Network and the Global Supernova Project. The LCO team is supported by U.S. NSF grants AST-1911225 and AST-1911151.
E. Malygin, D. Oparin, R. Uklein obtained part of the observed data on the unique scientific facility `Big Telescope Alt-azimuthal' of SAO RAS with the financial support of grant No075-15-2022-262 (13.MNPMU.21.0003) of the Ministry of Science and Higher Education of the Russian Federation.
Observations with the SAO RAS telescopes are supported by the Ministry of Science and Higher Education of the Russian Federation. The renovation of telescope equipment is currently provided within the national project `Science and Universities'.
This work was performed in part using equipment purchased with the support of the Moscow University Development Program.
Based on observations obtained at the 3.6\,m Devasthal Optical Telescope (DOT), which is a National Facility run and managed by Aryabhatta Research Institute of Observational Sciences (ARIES), an autonomous Institute under Department of Science and Technology, Government of India.
The LBT is an international collaboration among institutions in the United States, Italy, and Germany. LBT Corporation partners are The University of Arizona on behalf of the Arizona Board of Regents; Istituto Nazionale di Astrofisica, Italy; LBT Beteiligungsgesellschaft, Germany, representing the Max-Planck Society, The Leibniz Institute for Astrophysics Potsdam, and Heidelberg University; The Ohio State University, representing OSU, University of Notre Dame, University of Minnesota, and University of Virginia.
This publication makes use of data products from the Two Micron All Sky Survey, which is a joint project of the University of Massachusetts and the Infrared Processing and Analysis Center/California Institute of Technology, funded by NASA and the U.S. NSF.
This work has made use of data from the Asteroid Terrestrial-impact Last Alert System (ATLAS) project. The ATLAS project is primarily funded to search for near-Earth objects (NEOs) through NASA grants NN12AR55G, 80NSSC18K0284, and 80NSSC18K1575; byproducts of the NEO search include images and catalogs from the survey area. This work was partially funded by {\it Kepler/K2} grant J1944/80NSSC19K0112 and {\it HST} grant GO-15889, and STFC grants ST/T000198/1 and ST/S006109/1. The ATLAS science products have been made possible through the contributions of the University of Hawaii Institute for Astronomy, the Queen's University Belfast, STScI, the South African Astronomical Observatory, and The Millennium Institute of Astrophysics (MAS), Chile.
The Pan-STARRS1 Surveys (PS1) and the PS1 public science archive have been made possible through contributions by the Institute for Astronomy (the University of Hawaii), the Pan-STARRS Project Office, the Max-Planck Society and its participating institutes, the Max Planck Institute for Astronomy, Heidelberg, and the Max Planck Institute for Extraterrestrial Physics, Garching, the Johns Hopkins University, Durham University, the University of Edinburgh, the Queen's University Belfast, the Harvard-Smithsonian Center for Astrophysics, the Las Cumbres Observatory Global Telescope Network Incorporated, the National Central University of Taiwan, STScI, NASA under grant NNX08AR22G issued through the Planetary Science Division of the NASA Science Mission Directorate, U.S. NSF grant AST-1238877, the University of Maryland, Eotvos Lorand University (ELTE), the Los Alamos National Laboratory, and the Gordon and Betty Moore Foundation.
Pan-STARRS is a project of the Institute for Astronomy of the University of Hawaii, and is supported by the NASA SSO Near Earth Observation Program under grants 80NSSC18K0971, NNX14AM74G, NNX12AR65G, NNX13AQ47G, NNX08AR22G, 80NSSC21K1572, and 20-YORPD 20\_2-0014, as well as by the State of Hawaii.
{\sc iraf} was distributed by the National Optical Astronomy Observatory, which was managed by the Association of Universities for Research in Astronomy (AURA), Inc., under a cooperative agreement with the U.S. NSF.
This research has made use of the NASA/IPAC Extragalactic Database (NED), which is operated by the Jet Propulsion Laboratory, California Institute of Technology, under contract with NASA. 
This publication makes use of data products from the Near-Earth Object Wide-field Infrared Survey Explorer (NEOWISE), which is a joint project of the Jet Propulsion Laboratory/California Institute of Technology and the University of Arizona. NEOWISE is funded by NASA.

\end{appendix}


\end{document}